\documentclass[fleqn,usenatbib]{mnras}

\usepackage{newtxtext,newtxmath}

\usepackage[T1]{fontenc}

\DeclareRobustCommand{\VAN}[3]{#2}
\let\VANthebibliography\thebibliography
\def\thebibliography{\DeclareRobustCommand{\VAN}[3]{##3}\VANthebibliography}

\usepackage{graphicx}	
\usepackage{amsmath}	
\usepackage{nicematrix}
\usepackage{float}

\graphicspath{{images/}}

\DeclareMathOperator*{\argmin}{arg\,min}

\defcitealias{el2022unicorns}{E22}

\title[Autonomous Disentangling]{Autonomous Disentangling for Spectroscopic Surveys}

\author[Seeburger et al]{
Rhys Seeburger,$^{1}$\thanks{E-mail: seeburger@mpia.de}
Hans-Walter Rix,$^{1}$
Kareem El-Badry$^{1,2}$
Maosheng Xiang$^{3,4}$
Morgan Fouesneau$^{1}$
\\
$^{1}$Max Planck Institute for Astronomy, Heidelberg, Germany\\
$^{2}$California Institute of Technology, Pasadena, CA\\
$^{3}$National Astronomical Observatories, Chinese Academy of Sciences, Beijing, 100101, China\\
$^{4}$Institute for Frontiers in Astronomy and Astrophysics, Beijing Normal University, Beijing, 102206, China
}

\pubyear{2024}

\begin{document}
\label{firstpage}
\pagerange{\pageref{firstpage}--\pageref{lastpage}}
\maketitle

\begin{abstract}
A suite of spectroscopic surveys is producing vast sets of stellar spectra with the goal of advancing stellar physics and Galactic evolution by determining their basic physical properties. A substantial fraction of these stars are in binary systems, but almost all large-survey modeling pipelines treat them as single stars. For sets of multi-epoch spectra, spectral disentangling is a powerful technique to recover  or constrain the individual components' spectra of a multiple system. So far, this approach has focused on small samples or individual objects, usually with high resolution ($R \gtrsim 10.000$) spectra and many epochs ($\gtrsim 8$). Here, we present a disentangling implementation that accounts for several aspects of few-epoch spectra from large surveys: that vast sample sizes require automatic determination of starting guesses; that some of the most extensive spectroscopic surveys have a resolution of only $\approx 2,000$; that few epochs preclude unique orbit fitting; that one needs effective regularisation of the disentangled solution to ensure resulting spectra are smooth. We describe the implementation of this code and show with simulated spectra how well spectral recovery can work for hot and cool stars at $R \approx 2000$. Moreover, we verify the code on two established binary systems, the ``Unicorn'' and ``Giraffe''. This code can serve to explore new regimes in survey disentangling in search of massive stars with massive dark companions, e.g. the $\gtrsim 200,000$ hot stars of the SDSS-V survey.
\end{abstract}




\section{Introduction}
\label{sec:introduction}

The fact that a significant fraction of all stars or stellar remnants is in multiple stellar systems with a period of less than a few years \citep[e.g.][]{sana2012binary, moe2017mind} fundamentally affects many aspects of astrophysics. It affects the stellar evolution of both components, in some cases already during their main sequence phases, and more often in the evolved phases that will result in compact objects (such as white dwarfs, neutron stars, and black holes); it affects nucleosynthesis, the formation channels of supernovae, and the interpretation of photometric, astrometric, or spectroscopic sky surveys. And massive binaries -- or their descendants -- are the most prominent and frequent source of gravitational waves so far \citep[e.g.][]{abbott2023gwtc}.

Most of these systems cannot be spatially resolved, with projected separations that often are $\lesssim 1$~mas. However, their orbital velocities make it possible to separate the constituents of such multiple stellar systems in velocity space, especially if spectra at different orbital phases exist. We commonly categorize spectroscopic binaries into SB1 and SB2. Here, SB1 denotes a single-lined spectroscopic binary, where only one of the component spectra is apparent in the observations, and SB2 describes a double-lined spectroscopic binary, where two sets of lines are visible in the observed spectra. The approach of using multi-epoch observations of spectroscopic binaries to determine the components is called \textit{spectral disentangling} \citep[e.g.][]{bagnuolo1991tomographic, simon1994disentangling, hadrava1995orbital}.

In broad terms, spectral disentangling assumes that spectra of a presumed multiple stellar system -- when observed at different epochs -- can be described as the sum of two (or more) spectra that are invariant in their rest-frame, but whose radial velocities (RVs) change as a function time, reflecting orbital motion.  The mathematical foundation of spectral disentangling has been established for 30 years \citep[e.g.][]{bagnuolo1991tomographic, simon1994disentangling, hadrava1995orbital}. End-to-end disentangling requires the simultaneous, or iterative, solution to two problems, (a) reconstructing the rest-frame spectra of each component and (b) determining the components' radial velocities at each epoch or, alternatively, the orbital solution of the overall system. If the velocities at all epochs are known, the reconstruction of the disentangled spectra reduces to a linear $\chi^2$-optimisation problem, aiming to match the combined spectra at all the different epochs.

However, the application of spectral disentangling to large data sets has some serious practical limitations. First, some literature work has assumed that (a very good guess for) various system parameters can be obtained independently (e.g. \cite{ilijic2004thoughtscres}'s code CRES requires input of both the primary's and the secondary's velocities, shift-and-add as described in \cite{shenar2020hidden} and \cite{shenar2022tarantula} requires input of a few orbital parameters, see table \ref{tab:codes}). If the data are of limited resolution, or if the components contribute comparably to the total spectrum's absorption lines, this may not be possible. Second, there are several inherent degeneracies, foremost the fact that any featureless continuum portion of the spectrum can be assigned to either spectral component without consequences in the data match. Third, the approach works manifestly best in the regime of many epochs with data at high spectral resolution (compared to the orbital velocity changes) and at very high signal-to-noise, so that the dimmer component causes distinct changes in the combined spectrum. Finally, solving the full non-linear problem, i.e. optimising the model-data match over all possible primary velocities, mass ratios, and disentangled spectra is very time-consuming.

Over the last decade and for the next decade, vast spectral surveys are driving an exponential growth in the number of high-quality stellar spectra. Current or upcoming surveys include SDSS \citep{york2000sloan, kollmeier2019sdss}, LAMOST \citep{LAMOST}, DESI \citep{DESI}, WEAVE \citep{dalton2012weave} and 4MOST \citep{de20194most}. Further, the Gaia Data Release 4 will provide a vest set of spectra. Many of these surveys, in particular Gaia \citep{GaiaDR3} and SDSS-V \citep{kollmeier2019sdss} have multi-epoch observations scheduled across the entire sky. 

These surveys offer vast potential to map the stellar binary population via spectral disentangling. Some surveys, such as SDSS-V have explicit programs to systematically survey stars searching for massive dark companions; there spectral disentangling is crucial to identify ``contaminants'' with two luminous components \citep[e.g.][]{shenar2022tarantula, shenar2022x, mahy2022identifying}.

In this context of vast spectral surveys, new requirements -- or \emph{desiderata} -- arise for practical approaches to spectral disentangling.
The approach must be
\begin{itemize}
    \item fast, so that multi-epoch data for $10^4$ to $10^6$ systems can be analyzed.
    \item autonomous, in the sense that initial parameter guesses that permit sensible solutions must be found algorithmically and reliably.
    \item astrophysically flexible; many close binary systems will contain non-standard (stripped, accreting, rapidly rotating) stars.
    \item robust, given that large surveys typically have fewer epochs, less S/N and lower resolution than single-object studies
    \item user-friendly, as a wide community should be in a position to consistenly analyze different data sets, or reanalyze a given data sets with different constraints on acceptable solutions.
\end{itemize}

In this paper, we propose a new implementation of spectral disentangling, designed to address these issues. It wraps the process of finding starting guesses, solving for the disentangled spectra, and optimising the flux- and mass ratio parameters into one continuous pipeline written in Python. The approach is also fast enough that it can be applied to surveys of many thousands of objects. We include features such as regularisation to ensure desired properties in the spectra, as well as other adaptations on the original method to optimise the code for survey disentangling.

Due to Python's rise to popularity, we have elected to write this implementation of the method in this language. While at the surface level, speed might be a concern, many of Python's modules are partially written in compiled languages (such as C) and merely provide an interface familiar to the average Python user. Thus, this issue remains manageable. Python does have the advantage of being widely known in the scientific community, making an eventual release of the code as a package accessible to many.

\section{Spectral Disentangling Methodology}
\label{sec:methodology}

We first briefly review the existing method and codes in spectral disentangling in \ref{subsec:existing_codes}, then contrast disentangling as an approach with other methods in the literature in \ref{subsec:disentangling_vs_other_methods}. We lay out the desired characteristics of a code for survey disentangling in \ref{subsec:survey_disentangling}. In subsection \ref{subsec:prepocessing} we describe the prepocessing of the data. We explain the two-step process of optimising for the velocities and mass ratio in \ref{subsec:step_ii}, and how we obtain the systemic velocity and light ratio in \ref{subsec:com_velocity_and_light_ratio}.

\subsection{Established Disentangling Codes}
\label{subsec:existing_codes}

There are a number of existing codes, based on and building further upon the concepts introduced in \citet{bagnuolo1991tomographic},  \citet{simon1994disentangling} and \citet{hadrava1995orbital}. Some prominent examples of these include KOREL \citep{hadrava2004korel}, CRES \citep{ilijic2004thoughtscres}, fd3 \citep{ilijic2017fd3}, Spectangular \citep{sablowski2017spectral} and shift\&add \citep{shenar2022tarantula}. A brief summary of these codes can be found in table \ref{tab:codes}.

\begin{table*}

\begin{tabular}{llllll}
\multicolumn{6}{c}{\large{Comparison of prominent existing disentangling codes}}\\
\\

Code                   & KOREL           & fd3             & Spectangular        & CRES   & shift\&add \\
Author                 & \cite{hadrava2004korel}         & \cite{ilijic2017fd3}          & \cite{sablowski2017spectral}           & \cite{ilijic2004thoughtscres} & \cite{shenar2022tarantula} \\
Language               & Fortran         & C               & C++                 & C      & Python \\
WL or Fourier          & Fourier         & Fourier         & WL                  & WL     & WL (iterative) \\
Solves                 & Spectra + Orbit & Spectra + Orbit & Spectra + Orbit/RVs & Spectra & Spectra + K1, K2 \\
Third component?       & Y               & Y               & Y                   & N       & Y \\
Required Input         & Orbit           & Orbit           & Orbit/RVs           & RVs & $P$, $T_0$, $\omega$, $e$\\

\end{tabular}
\caption{This table summarises a number of properties of some prominent disentangling codes, which include the author(s) who wrote the codes, the programming language they are written in, whether disentangling takes place in wavelength or fourier space, whether they solve only for the spectra or also attempt to find the RVs and/or Orbital solutions, whether the code is equipped to handle a third component, and which input or guesses are required.}
\label{tab:codes}
\end{table*}

These codes employ one of two major methods for solving the \textcolor{blue}{linear} disentanging problem, either in fourier space (KOREL, fd3) or in wavelength space (CRES, Spectangular). More discussion on this can be found in e.g. \cite{ilijic2004obtaining}, but in summary, both methods come with their own advantages and disadvantages. The Fourier method, most notably, outperforms the wavelength-based alogorithm in terms of speed and thus allows for the implementation of further generalisations \citep{hadrava2009disentangling}. However, this comes at the cost of requiring all spectra to be sampled on the same grid, as well as giving each point the same weight, and implicitly assuming the resulting spectrum to be a periodic function of the wavelength \citep{sablowski2017spectral}. Performing the disentangling in wavelength space allows the user to extend the wavelength range on which the component spectra are computed based on the RV shifts of the individual epochs, as well as being less vulnerable to edge effects arising from the Fourier method.

\subsection{Disentangling vs Spectral Model Fitting}
\label{subsec:disentangling_vs_other_methods}

Spectral disentangling entails the ``non-parametric'' reconstruction of the individual components' rest-frame spectra. Alternatively, one can view the whole problem as a forward-modelling problem, drawing on a set of stellar templates. This has been explored and implemented by several groups \citep[e.g.][]{traven2020galah}.
An important downside of these methods is the fact that spectra in close binaries often do not look like simple, single-star spectra: they may be ``exotic'' objects such as stripped stars, they may show exceptionally fast rotation or may show emission lines from decretion disks. It can thus be easy to miss the signatures of ``strange'' companions \citep[e.g.][]{jayasinghe2021unicorn, jayasinghe2022giraffe, shenar2020hidden, el2022unicorns, bodensteiner2020hr, frost2022hr} when only considering a limited set of stellar templates.

This highlights the strength of template-independent methods, such as direct subtraction \citep{ferluga1997separating} iterative subtraction, also known as shift \& add \citep[][implemented in \citet{shenar2020hidden, shenar2022tarantula}]{gonzalez2006separation}, or, as described here, disentangling \citep{simon1994disentangling}. By not having to pre-select a template, we remain flexible to a range of potential outcomes of the procedure.

\subsection{A Disentangling Approach for Large Spectral Surveys}
\label{subsec:survey_disentangling}

\begin{figure*}[!h]
    \centering
    \includegraphics[width=\textwidth]{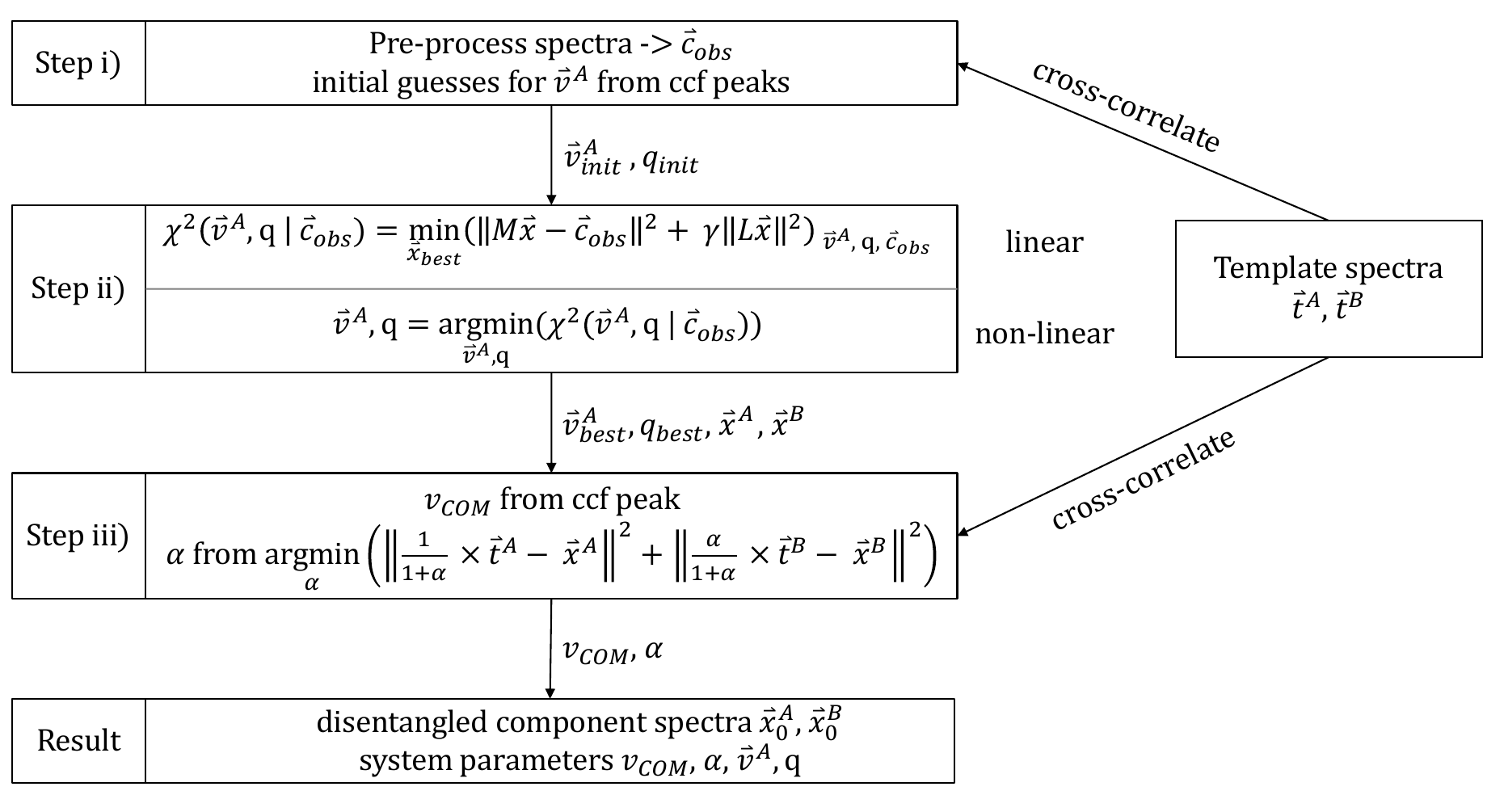}
    \caption{A flowchart showing the disentangling process as implemented in this work.}
    \label{fig:flowchart}
\end{figure*}

As mentioned in the Introduction, practical approaches to disentangling for vast spectral surveys call for a code that is fast, autonomous with respect to the initial parameter guesses, astrophysically flexible to accommodate the unusual spectra of close binaries, and robust with respect to suboptimal numbers of epochs, S/N and spectral resolution.

Here we set out to devise, test and verify such an approach. The end-to-end approach can be divided into several steps, which are conceptually illustrated in Figure \ref{fig:flowchart}, which serves as a guide and schematic representation of the process. 

We want to start with multi-epoch spectra for any given object, and at the end have both an estimate of the (usually two) disentangled spectra of the primary component's velocities at each epoch, the systemic velocity, the component mass ratio, and the mean component flux ratio. To achieve this for any one object, there are essentially three stages:
\begin{enumerate}
\item The spectra at all epochs for any one object have to be consistently pre-processed (normalisation, wevelength rebinning); and initial guesses for the velocities of the primary spectral component (defined as the component with the most prominent spectral lines) need to be derived algorithmically.
\item The optimisation of the main parameters consists of two parts. A non-linear part -- trying to find the best primary velocities at all epochs, $\vec{v}^A$ and the mass ratio $q$. And, a linear part, determining the two disentangled component spectra, $\vec{x}^A$ and $\vec{x}^B$ that best match the multi-epoch spectra $\vec{c}_{obs}$ for each assumed ($\vec{v}^A,q)$. 
\item The resulting spectra $\vec{x}^A$ and $\vec{x}^B$ have two remaining problems: they are pure mathematical constructs and know nothing about the rest-frame wavelengths of features. And any (modestly small) constant can be subtracted from $\vec{x}^A$ and added to $\vec{x}^B$, leaving the data match to the $\vec{c}_{obs}$. The last step then uses template spectra to a) fix the rest-frame of the disentangled spectra (or the systemic velocity of the system) and fix the luminosity ratio of the two components, by requiring physically sensible absorption line equivalent widths.
\end{enumerate}

We now describe these three stages in turn.

\subsection{Preprocessing of the Spectra}
\label{subsec:prepocessing}

Before disentangling, it proves convenient and useful to pre-process the observed multi-epoch spectra to simplify the subsequent math, labelled as step \emph{(i)} above. This first entails masking, or interpolating over,  bad pixels. Then resampling the spectra to a wavelength grid that is uniformly sampled in ln$\lambda$ (hereafter abbreviated as $\Lambda$), which linearises velocity shifts. Finally, we normalise the spectra by dividing them by a running median filter, where the filter must be chosen to be much wider than individual spectral features. The point of this is not necessarily to remove the ``continuum'' of the stellar spectrum, which is often conceptually and practically poorly defined, especially in cool stars with many spectral lines. We are much more concerned with removing the low-frequency variations in the observed spectra, as they may be dominated by instrumental effects. For our disentangling it is most important that we do this consistently across all observed epochs. We also apply the same median filtering to all the templates, i.e. we remove the low-frequency variations in the model spectra fully consistently. Other methods, such as polynomial fitting and clipping were considered, but ultimately median filtering was selected as the method of choice due to its speed, ease of use, and consistency across spectral types.

From this normalised spectrum, we then subtract 1 at all pixels, as this further simplifies the subsequent analysis. The different epochs are then concatenated into one long vector of length $N_{ep} \cdot N_{px}$ (number of epochs times number of pixels). We call this vector $\vec{c}_{obs}$. 

As the next part of pre-processing, we must find initial guesses for  the primary component's radial velocities for each epoch $j$ , $\vec{v}^A$), to aid the convergence of the non-linear parameter optimisation of step \emph{(ii)}.
We have implemented two ways to obtain these initial guesses, either using template cross-correlation, or using the TIRAVEL algorithm \citep[][described in the appendix]{zucker2006tiravel}.
For the cross-correlation, we first construct a grid of (rest-frame) template spectra from \citet{kurucz1979model}. These templates cover a suitable range of effective temperatures ($\sim 20$, 2.7kK to 25kK), surface gravities ($\sim 5$, -0.35 to 5.4), and two different rotation velocities $v$sin($i$) (10 and 100 km/s) and are matched in resolution and wavelength sampling to the observed spectra.  For any given epoch we perform a cross-correlation (CC) between the observed spectrum and all templates, and take the best template to be the one that yields the highest CC peak, when averaged over all epochs. The position of the CC peak for the best template yields our starting guess for the primary velocity at that epoch, yielding the initial $\vec{v}^A$.
As starting guesses for the mass ratio, we adopt $\frac{M_B}{M_A} \equiv q\equiv 1$.

In the system's center of mass frame, the primary and secondary velocities are related via $\vec{v}^B\equiv -q\times \vec{v}^A$. But the $\vec{v}^A$ we derive for each epoch are in the barycentric frame and we do not know the actual systemic velocity, $v_{COM}$. However, we can for the subsequent steps simply \emph{assume} that $v_{COM}\equiv 0$. This will then yield disentangled spectra of the correct shape, just in an ill-defined velocity reference frame. However, this can be remedied by correlation with template spectra in a subsequent step, as we show below.

\subsection{Parameter optimisation and initial disentangling}
\label{subsec:step_ii}

We can now proceed to step \emph{(ii)},  indicated in the second box (from the top) of the graphical representation in Figure~\ref{fig:flowchart}.
As mentioned, this step consists of two aspects. The linear part estimates the disentangled spectra that best match the observations at all epochs for given $\vec{v}^A$ and $q$. Each execution of this linear disentangling step yields $\chi^2(\vec{v}^A,q)$. This step is wrapped in a nonlinear parameter optimiser (we settled on the \cite{nelder1965simplex} method) that then finds most likely parameter estimates $\vec{v}^A_{best},q_{best}$ as

\begin{equation}
    \vec{v}^A_{best},q_{best}\equiv \argmin_{\vec{v}^A,q}\left(\chi^2(\vec{v}^A,q)\right).
\end{equation}

\subsubsection{The linear disentangling step}
\label{subsec:linear_matrix_algebra}

\begin{figure*}[H]
    \centering
    \includegraphics[width=\textwidth]{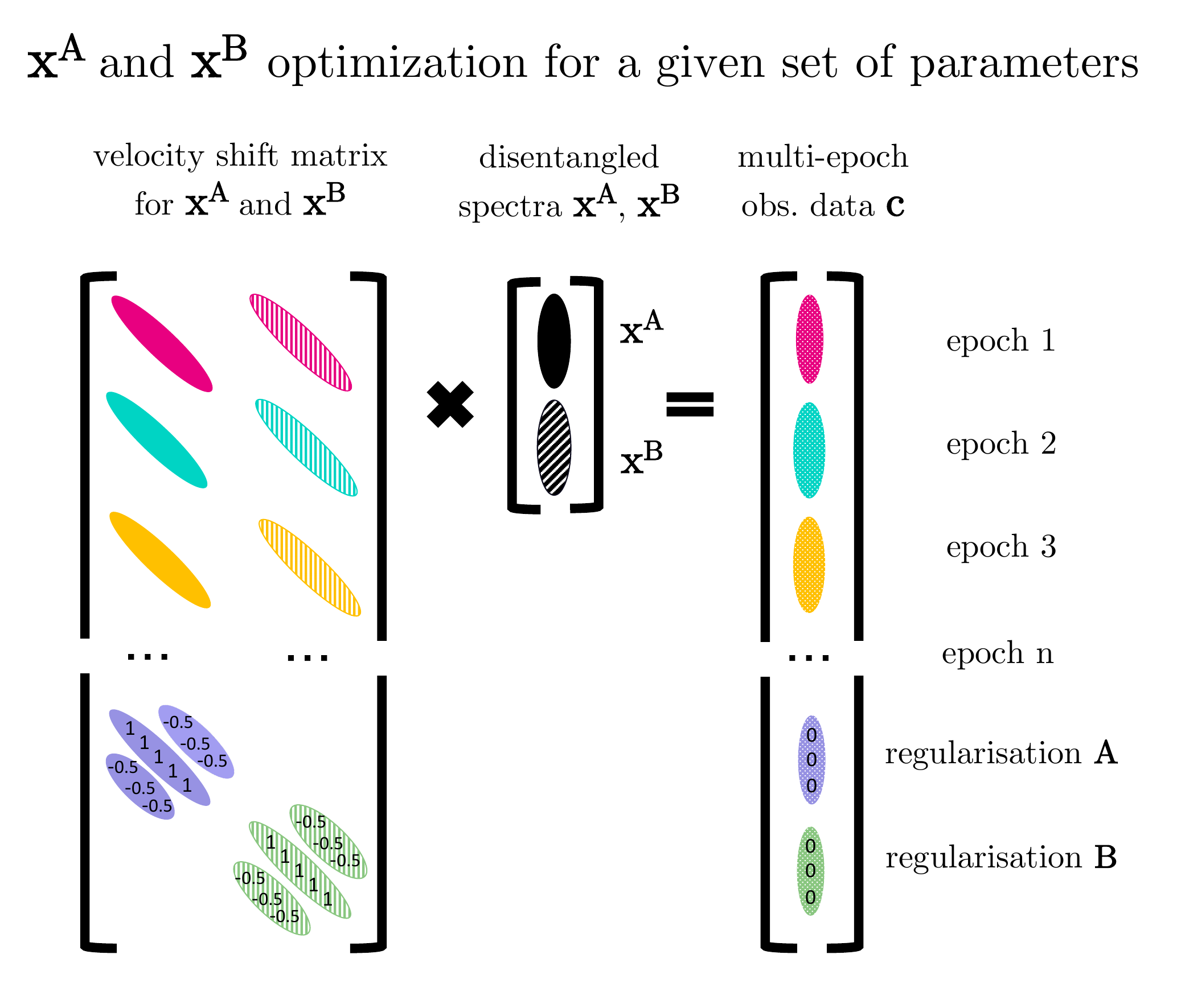}
    \caption{A cartoon of the setup of the linear algebra portion of the disentangling scheme. The top portion of matrix M on the left-hand side consists of (off-) diagonals whose position relates to the per-epoch velocity shifts. The vector in the center is simply the component spectra stacked on top of each other and the desired output of the linear algebra procedure. The vector on the right is the individual epoch composite spectra, stacked as well. The regularisation scheme detailed at the bottom of the cartoon is described in more detail in section \ref{subsec:regularisaiton}}
    \label{fig:schema}
\end{figure*}

It is more sensible to describe the linear disentangling step first. We restrict ourselves to the 
\emph{binary} (rather than triplet, etc.) case, where we seek the initially disentangled components $\vec{x}^A$ and $\vec{x}^B$, specified on a logarithmic wavelength grid with the same $\Delta \Lambda$ as the observed spectra: 
$\vec{x}^{A/B} \equiv \{x^{A/B}(\Lambda_{i})\} $ with $i=[0,N_{px}]$.

In this context, non-relativistic velocity changes correspond to constant shifts in $\ln{\lambda}$ and can be specified as
\begin{equation}
    \Delta \Lambda = \ln \left(1+\frac{v}{c}\right).
\end{equation}
In this first step we will retrieve 'initial' disentangled spectra in an ill-determined velocity frame rather than in the physical \emph{rest-frame}. We will remedy this in a subsequent step (Section~\ref{subsec:com_velocity_and_light_ratio}). For now, we can assume formally that the systemic velocity is not only constant but also zero; then the two component's velocity shifts at epoch $j$ are related via $\Delta \Lambda^B_{j} = -\Delta \Lambda^A_{j}/q $. In this case we only need to know the velocities for one component (say, A) and the system's mass ratio to determine all relevant velocities.


The shifted spectra of the components for each epoch $j$ can be written as:

\begin{align}
    \vec{x}^A_{j} &= \{x^A(\Lambda_i - \Delta \Lambda^A_{j})\} \\
    \vec{x}^B_{j} &= \{x^B(\Lambda_i - \Delta \Lambda^B_{j})\}.
\end{align}

The predicted composite spectrum $\vec{c}_{j, pred}$ at epoch $j$ is then given by:

\begin{equation}
    \vec{c}_{j, pred} = \{x^A(\Lambda_i - \Delta \Lambda^A_{j})+ x^B(\Lambda_i - \Delta \Lambda^B_{j})\}.
    \label{prematrix}
\end{equation}

The predictions for the composite spectra at all epochs $j$ can be cast more elegantly into a matrix of form:

\begin{equation}
    \mathbf{M} \cdot \vec{x} = \vec{c}_{pred},
\label{eq:Mxc}
\end{equation}

where the column vector $\vec{x}$ is the concatenation of $\vec{x}^A$ and $\vec{x}^B$, and $\vec{c}_{pred}$ the concatenation of all $\vec{c}_{j, pred}$ at all different epochs $N_{ep}$. Thus, $\vec{x}$ is a column vector of length $2 N_{px}$, and $\vec{c}_{pred}$ a column vector of length $N_{px} \cdot N_{ep}$. The matrix $\mathbf{M}$ is then a matrix of dimensions ($N_{px} \cdot N_{ep}) \times 2 N_{px}$ made of $2 \times N_{ep}$ blocks (one for each component and each epoch) of dimensions ($N_{px} \times N_{px}$). $\mathbf{M}$ is a sparse matrix, with the only nonzero elements being (off-)diagonals, whose position and value are governed by the per-epoch shifts of both components, $\Delta \Lambda^{A/B}_{j}$. Figure \ref{fig:schema} shows a simplified schematic of Equation~\ref{eq:Mxc}, and a more detailed description of the matrix structure can be found in Appendix \ref{app:matrix}.

For a given set of $\Delta \Lambda^A_{j}$ and q, we then wish to find the $\vec{x}_{best}$ for which $\vec{c}_{pred}$ best matches -- in the L2-norm or $chi^2$ sense -- the concatenated set of multi-epoch observations $\vec{c}_{obs}$:

\begin{equation}
    \vec{x}_{best} = \argmin_{\vec{x}} \lVert \mathbf{M} \cdot \vec{x} - \vec{c}_{obs}\rVert^2.
    \label{BIGBOY}
\end{equation}

As the matrix \textbf{M} is sparse and frequently very large, it makes sense to look for efficient methods for the use case. \cite{simon1994disentangling} propose the use of Singular Value Decomposition (SVD) \citep[e.g.][]{forsythe1977computer} to solve for $\vec{x}$. However, we found that the LSMR algorithm by \cite{fong2010lsmr} provided a more efficient iterative algorithm that exploits the sparsity of the matrix to arrive at a solution quickly. The resulting $\vec{x}_{best}$ can then be separated into $\vec{x}^A_{best}$ and $\vec{x}^B_{best}$.

\begin{figure*}
    \centering
    \includegraphics[width=\textwidth]{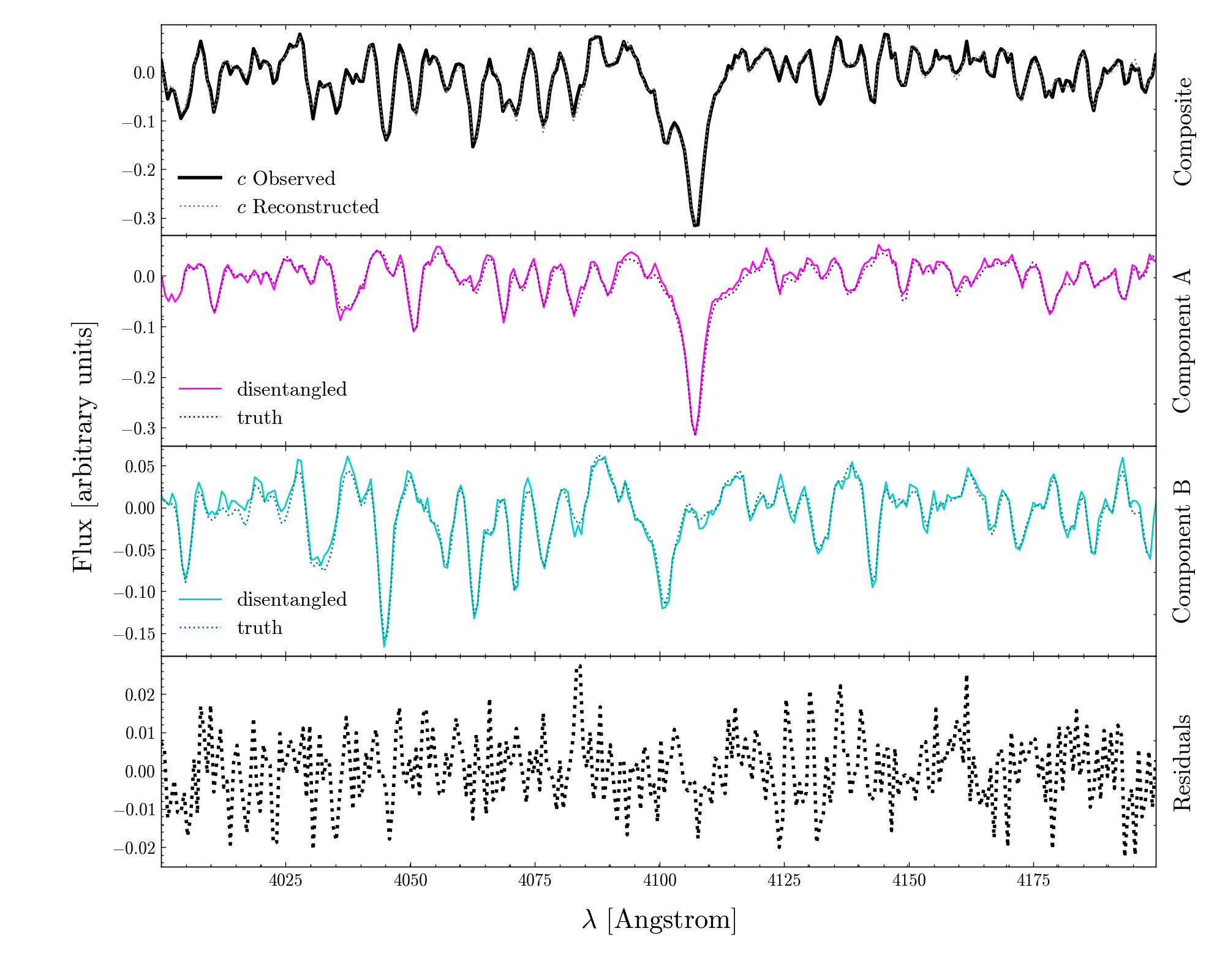}
    \caption{An example of a disentangled spectrum, displaying one epoch. The top panel shows the observed spectrum at that epoch in black, and the sum of the two disentangled and velocity-shifted component spectra in grey. The second and third panel show the rest-frame spectra for the disentangled components, with the true spectra shown as dotted lines. The bottom panel displays the difference between the observed and the reconstructed spectrum (shown separately in the top panel).}
    \label{fig:disentangled}
\end{figure*}

Figure \ref{fig:disentangled} shows an example of the disentangling result for a main-sequence binary consisting of a 1.1 $M_\odot$ primary with a 0.95 $M_\odot$ secondary, i.e. a mass ratio of $q = 0.86$. The velocity semi-amplitude of the primary is 200 km/s, we have assumed a circular, edge-on orbit and sampled the velocities regularly for 6 epochs at a resolution of $R = 2000$ and with a signal-to-noise ratio of 30. The systemic velocity is 100 km/s. Disentangling has been performed assuming $\Delta \Lambda^A_{j}$ and $q$ are known. Both the composite spectra (black and grey lines) and the individual disentangled component spectra are in excellent agreement with the observations (top) and input spectra (two panels below).

\subsubsection{Regularisation}
\label{subsec:regularisaiton}
While the method described so far can yield acceptable results, the two disentangled spectra are often not smooth. This is due to the nonuniqueness of the solutions: especially for sections of the continuum with no/few lines, it is ill-defined how much each component spectrum is specifically contributing to the composite.

Tikhonov regularisation \citep{tikhonov1963solution, phillips1962technique} introduces an additional term in the least-squares minimisation (Equation~\ref{BIGBOY}), which can be used to penalise certain undesired features in the solution, and encourage desired properties. So, instead of minimising the expression in Equation \ref{BIGBOY}, we seek to minimise:

\begin{equation}
\lVert \mathbf{M} \cdot \vec{x} - \vec{c}_{obs} \rVert^2 - \gamma \lVert \mathbf{L} \cdot \vec{x}\rVert^2,
\end{equation}

where \textbf{M}, $\vec{x}$ and $\vec{c}_{obs}$ are as previously described. \textbf{L} is a matrix by which we regularise the solution for $\vec{x}$ and $\gamma$ expresses the weight of this regularisation compared to the disentangling problem. There is a trade-off: very small $\gamma$ will lead to very little regularisation (and $\gamma = 0$ reduces to the original problem posed), while very large $\gamma$ will over-regularise the solution, leading to a loss of the original features.

This can be recast into a form very similar to Eq.\ref{BIGBOY}:

\begin{equation}
 \vec{x}_{best} = \argmin_{\vec{x}}
\left\|\left[\begin{array}{c}\mathbf{M} \\ \sqrt{\gamma} \mathbf{L}
\end{array}\right] \vec{x} -\left[\begin{array}{c}
\vec{c}_{obs}\\
\vec{0}
\end{array}\right]\right\|^2.
\label{eq:regularised_linear_opimisation}
\end{equation}

Here, \textbf{L} is of shape ($2 N_{px} \times 2 N_{px}$), and $\vec{0}$ is of length $2 N_{px}$. Thus, this extension to the original problem allows regularisation of $\vec{x}$ by simply appending to the matrix \textbf{M} and the vector $\vec{c}_{obs}$.

We want matrix \textbf{L} push for smooth $\vec{x}$ solutions but without enforcing a specific shape onto the spectra. We do this by minimising the second derivative or curvature among adjacent elements of $\vec{x}_{best}$ .  In practice, the matrix \textbf{L} has $1$'s along the diagonal , and  -1/2 along the two first off-diagonals above and below the main, as illustrated in the scheme shown in Figure~\ref{fig:schema}. This matrix leads to a norm that is simply the sum of the second derivatives among all sets of adjacent 3 pixels (expressed as a finite difference). Therefore, it penalizes strong local curvature of the result, and thus ``jaggedness'' of the disentangled spectra.

\begin{figure*}
    \centering
    \includegraphics[width=\textwidth]{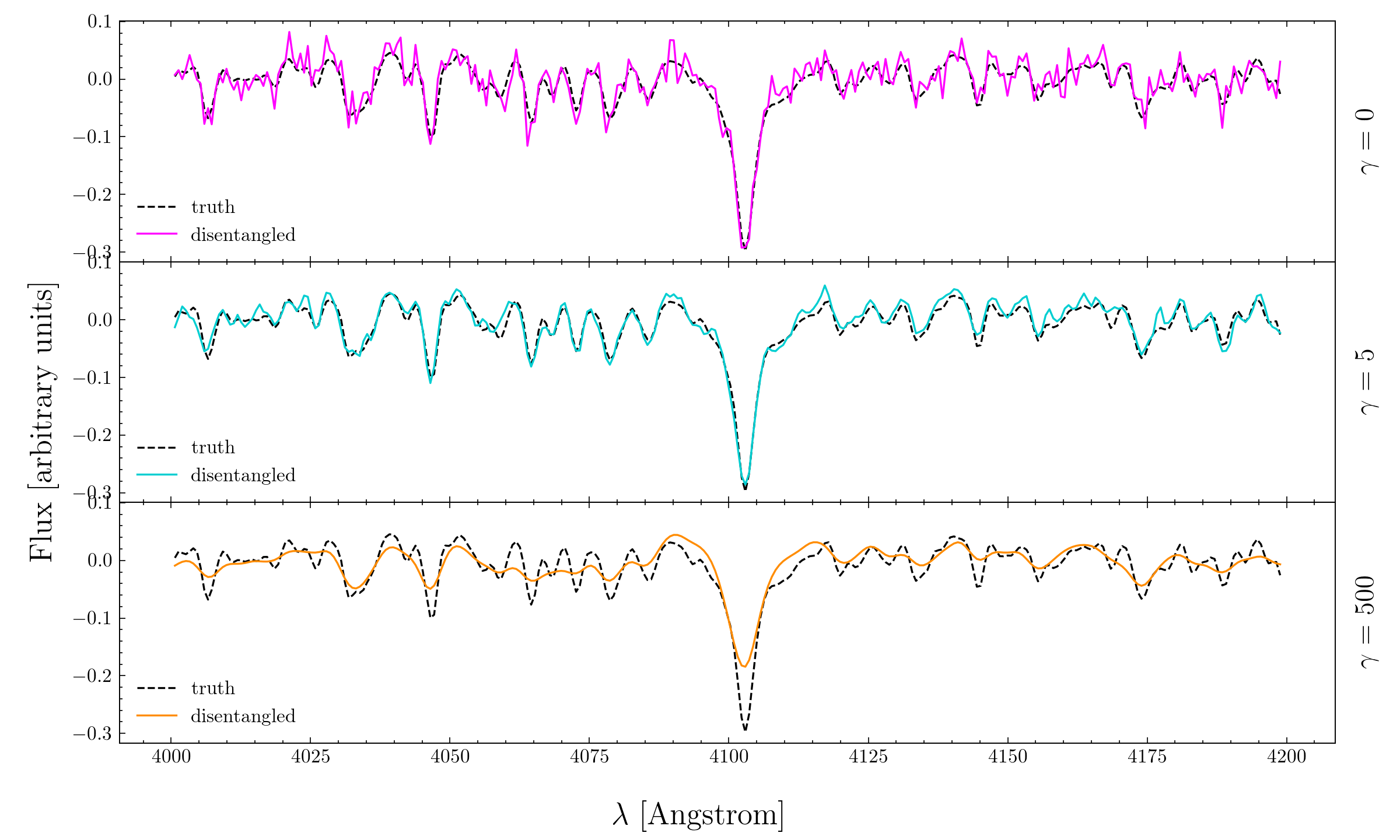}
    \caption{The spectrum of the primary as obtained from disentangling with varying degrees of regularisation. The first panel shows the result without regularisation, the second panel with $\gamma = 5$ and the last with $\gamma = 500$. In each panel, the true spectrum is shown with a black dashed line.}
    \label{fig:regularisation}
\end{figure*}

In Figure~\ref{fig:regularisation} we compare the result of different regularisation strengths, ranging from none (top) to a good amount (middle) to too much (bottom). We see that when no regularisation is applied, the disentangled spectrum is jagged due to the degeneracies. With an appropriate amount of regularisation, this can be remedied. If too much regularisation is applied, features and details will start to be ``washed out'' - it is thus important to choose the regularisation parameter correctly, e.g. by requiring it not increase the least-squares or $\chi^2$ significantly. In practice, we seek a regularised solution that fits the data to within $\Delta \chi^2 \sim 3$ as well as the unregularised solution. We have found a suitable $\gamma$ for each (simulated) dataset by trial and error; automatisation of this is an endeavour for future work.

\subsubsection{The non-linear optimisation step}
\label{subsec:non-linear_optimisation}

So far, we have shown how to robustly solve the disentangling problem if the RVs $\vec{v}^A$ for all epochs ${j}$ and the mass ratio $q$ are known.

We now need to implement a robust way to find the optimum parameters $\vec{v}^A$  and $q$ such that the two constructed component spectra $\vec{x}^A$ and $\vec{x}^B$, shifted by their respective epoch velocities, optimally reconstruct the observed composite spectrum $\vec{c}_{obs}$ for all epochs. To this end, we define the residual between the observed and reconstructed spectra as:

\begin{align}
    \vec{R} &=  \vec{c}_{j, obs} - \vec{c}_{j, pred}
\end{align}

or, more explicitly,

\begin{align}
    \vec{R} &=  \vec{c}_{j, obs} - \left( \vec{x}^A(\Lambda_i - \Delta \Lambda^A_{j})+ \vec{x}^B(\Lambda_i - \Delta \Lambda^B_{j}) \right).
    \label{eq:residual_with_explicit_lambda_dependence}
\end{align}

Equation \ref{eq:residual_with_explicit_lambda_dependence} explicitly recognises that the residuals are a function of $\Delta \Lambda^{A/B}_{j}$.  As the next step, we have to find the best values for these $\Delta \Lambda^{A/B}_{j}$ by minimising the value of $\chi^2$ as a function of $q$ and $\vec{v}^A$, defined as:

\begin{equation}
    \chi^2(\vec{v}^A, q \ | \ \vec{c}_{obs} ) = \vec{R}^T \cdot \mathrm{cov} \cdot \vec{R},
\label{eq:residual}
\end{equation}
where $\mathrm{cov}$ is the covariance matrix of the observed spectra.

We do this minimisation via the Downhill Simplex Optimisation by \cite{nelder1965simplex}, specifically the \verb|scipy.optimise | implementation. 

Robust optimisation requires a sensible initial starting guess.
We initialise it with $\vec{v}^A$ and $q$ as described in \ref{subsec:prepocessing}. Then the optimiser solves at each step the linear problem (Equation~\ref{eq:regularised_linear_opimisation}) for the current values of $\vec{v}^A$ and $q$. This returns the component spectra $\vec{x}^A$ and $\vec{x}^B$ and from them $\vec{c}_{pred}$, which yields $\chi^2$ from Equation~\ref{eq:residual} from the comparison with the data $\vec{c}_{obs}$. Eventually, we expect the optimiser to return the best values for the RVs and mass ratio, those that allowed for the most accurate reconstruction of the observed spectra. 

\subsection{Determining the systemic velocity and light ratio}
\label{subsec:com_velocity_and_light_ratio}

Mathematically, spectrum disentangling is possible while only knowing the relative per-epoch RVs of the primary and the mass ratio, as has been demonstrated. However, this yields two disentangled spectra that are in an unspecified velocity frame that is neither the center-of-mass frame nor the  rest-frame; and the velocity frames for the two components will generally not be the same. But in practice, we are interested in the systemic velocity $v_{COM}$ and the light ratio of the two components $\alpha$.

We now describe how we find these parameters and move the disentangled spectra to a well-defined rest-frame, as outlined in step 
\emph{(iii)} of Figure \ref{fig:flowchart}. In doing so, we basically draw on 
external astrophysical information: the rest-frame wavelengths and (approximate) equivalent widths of prominent stellar absorption lines in the disentangled spectra must resemble -- at least roughly -- those in a comprehensive set of template spectra.

We start by using such template spectra to find the systemic velocity. We do this by computing the cross-correlation (CC) of the initial solution vectors $\vec{x}^{A/B}_{best}$ with spectra in the template set and identifing the best template. We denote the velocity of that CC peak as $\epsilon^{A/B}$, which describes the offset of the velocity found by the optimiser from the true observed velocity of the primary. In this notation we have

\begin{equation}
    \Tilde{v}^A = v^A + \epsilon^A = w^A + v_{COM} + \epsilon^A
    \label{eq:offset_0}
\end{equation}

where $\Tilde{v}^{A}$ are the velocities in the initial ill-defined frame, $v^A$ the true velocities in the observer frame, and $w^{A}$ are the center-of-mass-frame velocities. With starting guesses via template cross-correlation, we expect $\epsilon^A$ to be close to zero. This is because the templates are in the rest frame, so initial velocities found using them should also be in the rest frame, meaning the offset $\epsilon^A$ should be small. TIRAVEL \citep{zucker2006tiravel} determines initial guesses by cross-correlating the different epochs with each other and solving for the most probable vector of RVs. Then, $\epsilon^A$ could be large, as we have no prior information about absolute RVs, only relative (to other epochs).  By contruction from the disentangling, the primary and secondary velocities in our ill-defined frame  are related by $\Tilde{v}^B = -\Tilde{v}^A/q $ holds. And, we have the relations

\begin{equation}
        v^B = w^B + v_{COM} = -\frac{w^A}{q} + v_{COM},
        \label{eq:offset_1}
\end{equation}

with nomenclature analogous to Equation \ref{eq:offset_0}; Eq.~\ref{eq:offset_1}  simply states that the true observed velocity is the secondary's center-of-mass frame velocity plus the systemic velocity, which in turn is related to the true velocity of the primary via the mass ratio $q$.
We can now use these offsets $\epsilon^{A/B}$ and Equations \ref{eq:offset_0} and all \ref{eq:offset_1} to determine the systemic velocity of the system:

\begin{align}
    \epsilon^B &= \Tilde{v}^B - v^B \ \  = \ \ -\frac{\Tilde{v}^A}{q} +\frac{w^A}{q} - v_{COM}  \\
     v_{COM} &= - \frac{\epsilon^B + \epsilon^A / q}{1 + 1 / q}
    \label{eq:offset_2}
\end{align}

Note that even when $\epsilon^A$ is small, $\epsilon^B$ might be large, especially when the originally assumed systemic velocity is far from the truth. 

This puts us in a position to solve for the component spectra in the rest-frame using these results and Equation~\ref{eq:offset_0}. This process also gives us templates $\Vec{t}^A$ and $\Vec{t}^B$, the templates that had the highest CC peak with the respective disentangled component spectra.

We can now turn to determining the light ratio $\alpha$, using the best-fitting templates. In the end we attribute different portions of the ``featureless continuum'' to the two disentangled  components so that their equivalent widths are physically plausible. We obtain $\alpha$ by minimising

\begin{equation}
    \argmin_{\alpha} \lVert  \frac{1}{1+\alpha} \Vec{t}^A - \vec{x}^A + \frac{\alpha}{1+\alpha} \Vec{t}^B - \vec{x}^B \rVert^2,
\end{equation}

which scales the templates for both components, $\Vec{t}^A$ and $\Vec{t}^B$, such that the scaled templates most resemble the disentangled spectra.

\begin{figure}
    \centering
    \includegraphics[width=0.45\textwidth]{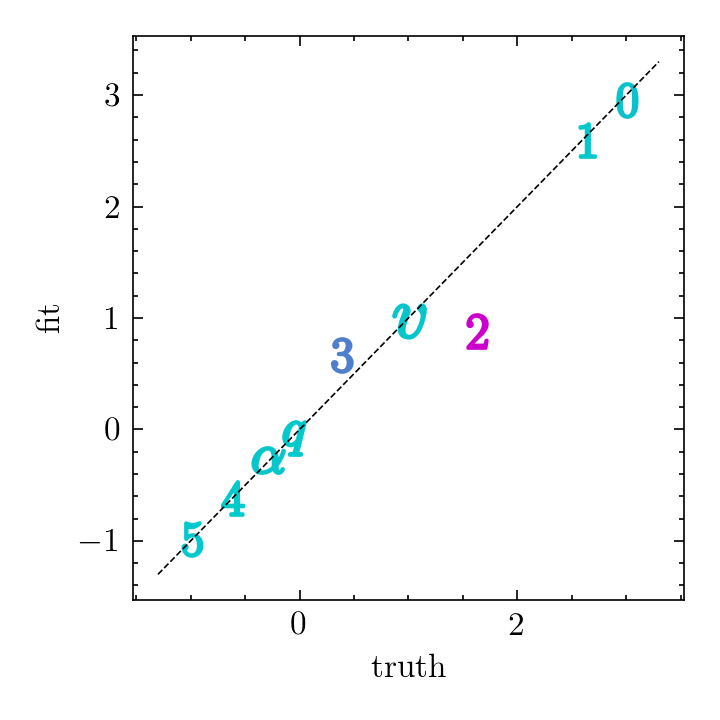}
    \caption{An example of the results for the different parameters obtained by the optimiser. The correct value is plotted along the x-axis, with the value recovered by the optimiser along the y-axis. The dashed line indicates agreement between the two, and the further away from this line a value falls, the less accurately it was determined. The colour of each symbol also indicates how well the parameter was fit, with pink indicating a bad, and cyan a good agreement of the fit value with the truth relative to the other parameters. The symbol used for each point represents the parameter, with numbers 0-5 indicating the per-epoch RVs for each epoch (divided by 100), $q$ the $log_{10}$ of the mass ratio, $v$ the systemic velocity (divided by 100) and $\alpha$ the $log_{10}$ of the light ratio. We selected these scaling of the different parameters for better visualisation. We see good recovery for all parameters, except two of the RVs. This is, however, not further surprising, as these RVs are fairly close to the systemic velocity, meaning there is only a small doppler shift of the two spectra relative to each other, making determination of the accurate velocities difficult at these low resolutions.}
    \label{fig:fit}
\end{figure}

An example of the results of this 3-step (preprocessing, optimising $\vec{v}^A$ and $q$, finding $v_{COM}$ and $\alpha$) process can be seen 
in Figure \ref{fig:fit}, where the accuracy of each parameter as found in the optimisation or subsequent step, is explored. The system has the same parameters as the one discussed in section \ref{subsec:linear_matrix_algebra} whose disentangled spectrum can be seen in Figure \ref{fig:disentangled}.

\section{Algorithm Validation}
\label{sec:algorithm_validation}

We now proceed to validate this autonomous multistep approach to spectral disentangling that we have laid out. In particular, we want to explore under which circumstances it yields sensibly disentangled spectra and reasonable physical parameters. This will depend on both the physical properties of the binary system (velocity semi-amplitude $K$, mass and light ratios, effective temperatures) and on the observational set-up (S/N, spectral resolution, number of epochs).

In exploring this, we will particularly focus on the observational parameter regime pertinent to large spectroscopic surveys: modest resolution and S/N along with relatively few epochs. We do this by simulating composite spectra of binaries with both hot and cool primaries, and then running the algorithm 'blindly', or autonomously,  on them.
These simulations allow us to assess both the \emph{robust} and the \emph{problematic} disentangling  regimes. After these simulations, we illustrate the approach using a well-established binary system.

\subsection{Simulated Data}
\label{subsec:simulated_data}

We perform an initial test of the method on simulated data, given that it allows us to precisely control the system parameters and check how well the algorithm recovers them. To simulate co-evolving binary components, we use a 1 Gyr and 250 Myr isochrone from \citet{bressan2012parsec} for the cool- and hot-star primary simulations, respectively, selecting a 1.1 $M_\odot$ and 3 $M_\odot$ star as the primary.

For resolutions of $R \approx 2,000$ and $R \approx 20,000$, we then explore a grid of different RV semi-amplitudes $K$, linearly spaced from 50 km/s to 250 km/s, and light ratios $\alpha$. For $\alpha$, we consider logarithmically spaced values from 0.01 to 1 to explore the more ``extreme'' regimes of very faint secondaries, as well as a linearly spaced grid from 0.1 to 0.9. With the primary's parameters (mass, age,  $T_{\rm eff}$ and $log(g)$) and $\alpha$, we can use the isochrones to get analogous parameters for the secondary; the two components' masses then also yield $q$. We then select spectral templates, using  \cite{kurucz1979model} spectra for the two components. Then, we set the orbital parameters of the system (assuming a circular, edge-on orbit), sample the RVs of both components at 6 different epochs uniformly in phase-space over half an orbit, and create composite spectra, adding noise. Here, we assumed a signal-to-noise ratio of 30.

We feed these simulated spectra to the disentangling and optimisation algorithm, assessing how well it is able to obtain the correct system parameters, reconstruct the observed composite spectra $\vec{c}_{obs}$ and solve disentangled rest-frame spectra $\Vec{x}^A_{0}$ and $\Vec{x}^B_{0}$.

We assess the quality of the parameter through this figure of merit:

\begin{equation}
    \rm{FOM} = 1 - \frac{1}{n} \sum_{i=1}^{n} w_i \cdot \frac{|\theta_{i, pred} - \theta_{i, true}|}{|\theta_{i, pred}| + |\theta_{i, true}|},
\end{equation}

which is bound between 0 and 1, where $\theta_{i, \rm 
 pred}$ are the estimates returned by our algorithm,  $\theta_{i, \rm  true}$ the true (simulation input) values, and $w_i$ are the normalised weights assigned to each parameter. We have assigned weights of 0.25 to $q$, $v_{COM}$ and $\alpha$, and a weight of $\frac{0.25}{N_{ep}}$ to each of the primary RVs. The weighting avoids the FOM being dominated by the accuracy of the recovered primary RVs (especially in the case of many epochs), which are comparatively easy to find, and places greater importance in correctly determining the parameters associated with the secondary. It is pertinent to note, at this point, that a ``good'' FOM means that the system parameters have been recovered accurately, not necessarily that the disentangled spectra are equivalent to the ``ground truth''. It is simply a metric to assess the performance of the optimiser.

\begin{figure*}
    \centering
    \includegraphics[width=\textwidth]{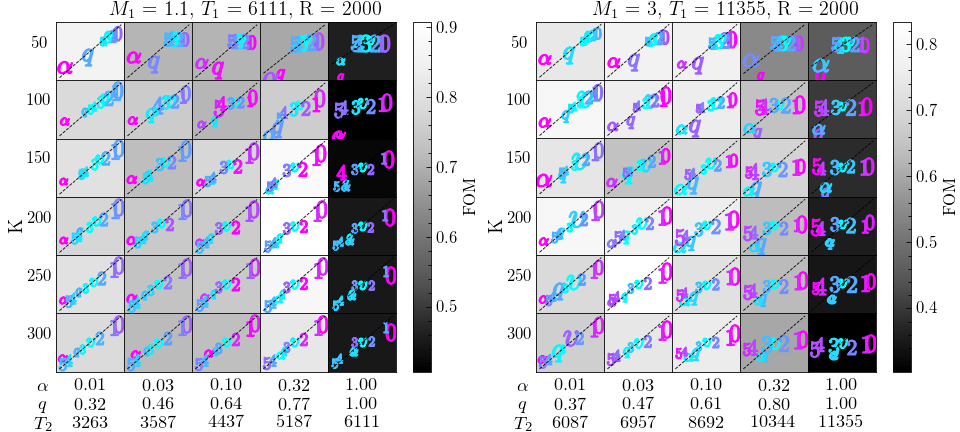}
    \caption{A grid of the results of the autonomous disentangler on a range of different systems. The left panel shows the case of a low mass primary (1.1 $M_\odot$), while the right panel displays the case of a  higher mass primary (3 $M_\odot$). Each panel consists of multiple cells, varying in the semi-amplitude of the radial velocities, $K$, along the y-axis, and in light ratio $\alpha$, and consequently, the mass ratio $q$ and effective temperature of the secondary, $T_{\rm eff, 2}$, along the x-axis.} The colour of each cell indicates the figure-of-merit value the disentangler achieved for the recovery of the parameters, with white indicating a higher, and thus better, figure-of-merit, and black indicating a lower, worse FOM. Each cell also contains the accuracy and precision achieved for each individual parameter, analogous to Figure \ref{fig:fit}. The x-axis here displays the true value of the parameter, and the y-axis the one recovered by the optimiser. If the optimiser determined the correct parameter, we expect it to lie along the dashed grey diagonal line. The colour of each parameter indicates how well it was recovered within its specific system, with pink indicating a comparatively large distance between truth and fitted parameter, and cyan indicating a small distance, and thus good agreement. Lastly, the size of each symbol is related to its variance in the bootstrapping process, with a large symbol indicating a large variance (and thus a low precision) and a small symbol indicating a small variance and higher precision. 
    \label{fig:grid}
\end{figure*}

\begin{figure*}
    \centering
    \includegraphics[width=\textwidth]{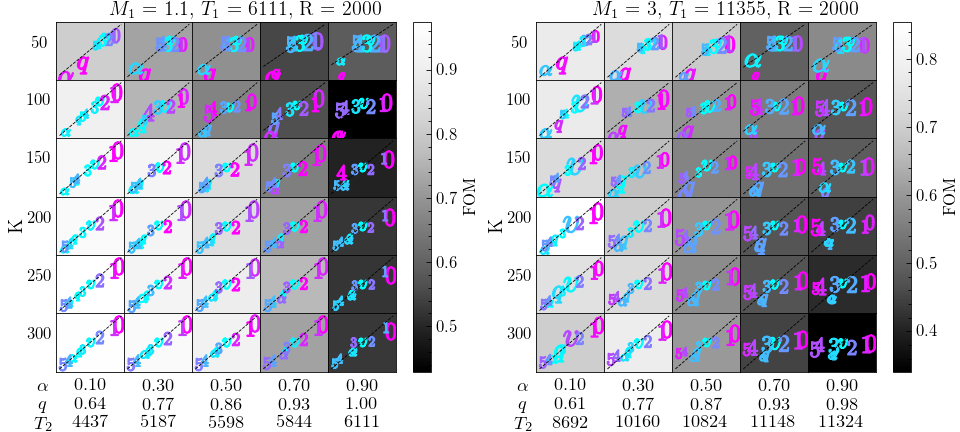}
    \caption{Analogous to Figure \ref{fig:grid}, with the light ratio now varying between 0.1 and 0.9 in steps of 0.2.}
    \label{fig:grid_small}
\end{figure*}

The results of this validation are shown in Figure \ref{fig:grid} for cool (left panel) and hot primaries (right panel). The different subpanels show a grid in velocity semi-amplitude $K$ and light ration $\alpha$. The background color of each subpanel indicates the FOM, with lighter colors indicating better parameter retrieval. For high velocity semi-amplitudes (compared to the spectral resolution) and distinctly different component temperatures (whenever $\alpha\ll 1$) the parameter retrieval is robust and quite precise; this is particularly true if the primary star is already cool. But the Figure also shows that there are two regimes where our algorithm struggles: first, for a mass ratio of unity (twin star spectra) there is a degeneracy between the two components' velocities. Consequently, the optimiser cannot determine the other parameters correctly. For the very faint secondary regime ($\alpha = 0.01$), the optimiser finds the correct primary velocities, but has some difficulties with the mass and light ratio, as well as q.  This is not particularly surprising, as in this regime the signal of the secondary is on the level of the noise, and thus finding parameters that pertain not just to the primary (as the RVs do, in this case) is problematic.

Second, Fig.~\ref{fig:grid} also shows  that lower velocity semi-amplitudes (below the spectral resolution) lead to a worse determination of the individual component's velocities, and consequently of the other parameters. For the higher-mass and hotter primary, the broader spectral lines and comparatively similar spectra between the primary and secondary also lead to an underestimation of the RVs, as the CFF peak is substantially affected by the secondary.

Figure \ref{fig:grid_small} is analogous to Figure~\ref{fig:grid}, with $\alpha$ varying from 0.1 to 0.9 in steps of 0.2. Again we see that for both the cool and the hotter primary, our disentangler has difficulties correctly recovering the velocities for the lowest RV semi-amplitudes. As we are now excluding the more extreme cases of twin stars and extreme light ratios, where the secondary contribution is on the level of the noise, we observe generally higher (and thus better) figure-of-merit values. With the linear grid in $\alpha$ we can see a more gradual trend in the robustness of our disentangling approach, with accuracies getting worse as we get ``too close'' to an equal mass binary, especially for the 3M$_\odot$ primary, as the spectra of the primary and secondary show rather similar features.

\begin{figure*}
    \centering
    \includegraphics[width=\textwidth]{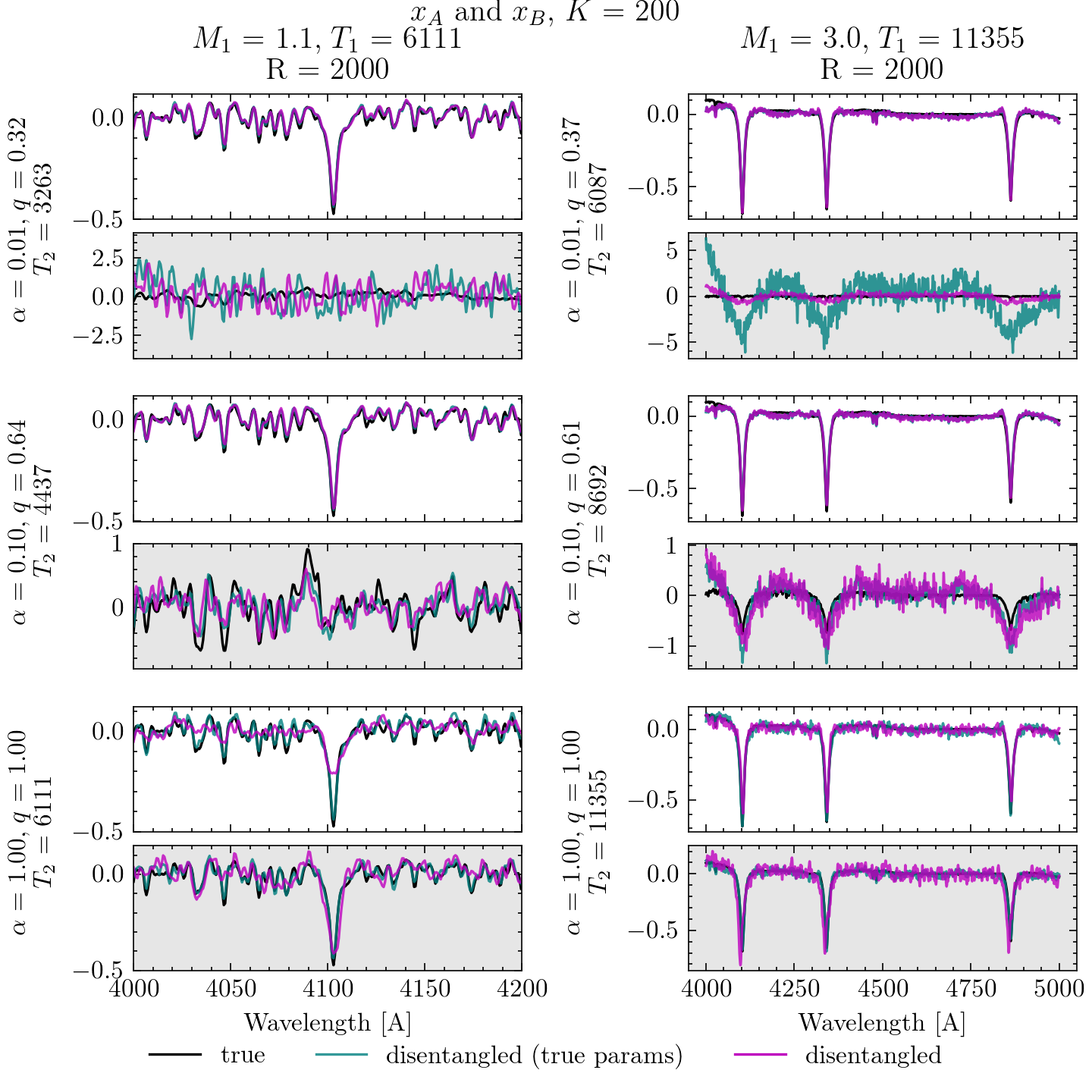}
    \caption{A selection of individual results of the autonomous disentangler, here for a RV semi-amplitude of 200 km/s. The left column shows the low-mass primary case, and the right the higher-mass primary. The different rows show different light ratios $\alpha$ and thus mass ratios $q$ and effective temperature of the secondary, $T_{\rm eff, 2}$. The top panel (white background) in each row shows the result of the disentangling for the primary, and the bottom (grey background) for the secondary. For each panel, the truth is plotted in black, with the disentangling result achieved by using the ``correct'' input parameters in green, and the actual result cyan the autonomous disentangler in magenta. The disentangler solutions are scaled by the light ratio (true value for cyan, value recovered by the optimiser for magenta) to match them to the true input. Thus, a ``correctly shaped'' spectrum that is scaled incorrectly indicates an incorrect light ratio, while a spectrum that is offset relative to the truth indicates that the systemic velocity was not found correctly. For the most extreme light ratios (top row), even with ``correct'' input parameters, the disentangler struggles to recover the correct spectrum of the secondary. The optimiser also has difficulties recovering the correct spectrum in the higher-mass case, owing to the wide lines found in hot stars.}
    \label{fig:disentangled_all}
\end{figure*}

Figure \ref{fig:disentangled_all} shows individual disentangled spectra for different simulated systems. We  show a smaller wavelength range for the cooler primary, as there are more and narrower lines, whereas for the hotter primary, the lines are wider and fewer. We see that the disentangler successfully recovers the primary component in all cases, using both the ``true'' parameters as well as the ones found by the optimiser. For the smallest light ratio of 0.01, both in the hot- and the cool-star case, the disentangler struggles to recover the secondary, due to its small contribution to the spectrum. There are still some issues with the secondary spectrum recovery for the hotter primary even at a more moderate light ratio of 0.1. We believe this to be due to the very similar lines between the primary and secondary - for moderate velocity shifts, two similar spectra being shifted against each other look like the lines are ``widening'' and ``narrowing'', rather than fully seperating, which causes some issues in the disentangling process. 

\subsubsection{Disentangling SB1 Systems with Dark Companions}
Part of the motivation for this work is to find systems where the secondary is dark, i.e. does not show up in the disentangling. To explore how our code responds to such systems, we have simulated a binary consisting of a 1.1 $M_{\odot}$ primary with a dark companion (a so-called SB1 system). Like before, we set a velocity semi-amplitude of 200 km/s for the bright component and assume a circular, edge-on orbit. RVs are sampled uniformly in time at a spectral resolution of $R = 2000$ and a signal-to-noise ratio of 30. 

We then applied our optimiser and disentangling procedure to this system in the same way as the previous simulations, presuming we have no prior knowledge of its SB1 nature. Inevitably, the optimiser will find some $v_{COM}$ and $q$, which must be spurious. And then it determines an $\alpha$ from the disentangled spectra that were computed using these $v_{COM}$ and $q$. We would expect the secondary spectrum to be essentially noise, and the resulting $\alpha$ to be small, presumably an upper limit.

\begin{figure}
    \centering
    \includegraphics[width=0.45\textwidth]{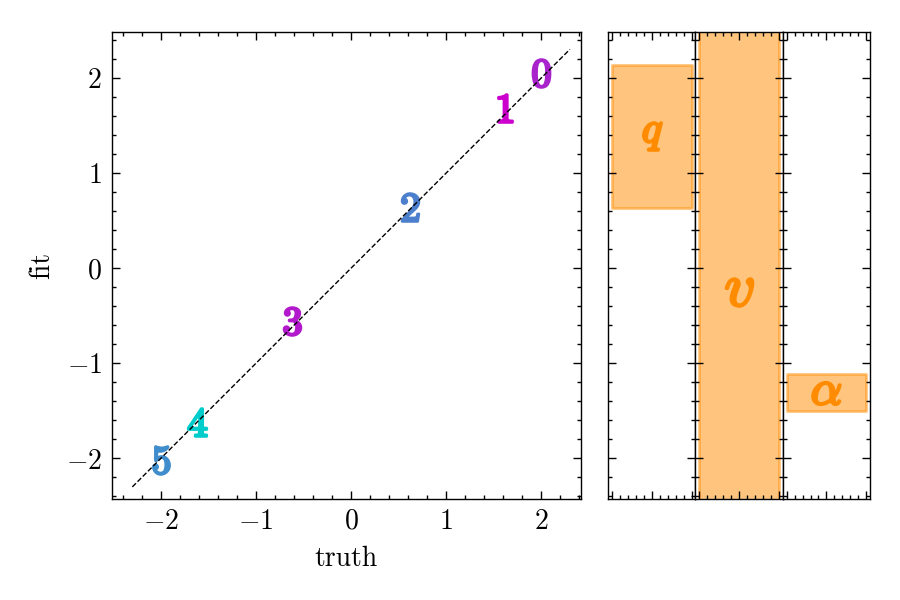}
    \caption{Formal disentangling results for a simulated SB1 system (dark companion), analogous to figure \ref{fig:fit}.
    The single-epoch velocities of the luminous component are well-determined. The parameters $v_{COM}$ and $q$ are ill-determined (because there is no second spectrum); the 1 $sigma$ range for one noise-realization is shown by the orange bands in the right panels. The value for $\alpha$ in such cases (implicitly an upper limit) is always found to be small.}
    \label{fig:fit_a0}
\end{figure}

Figure \ref{fig:fit_a0} shows the results of the optimiser in one realization of this scenario. We see that the primary velocities are found near perfectly. In the right panels, we show the formal results for the three problematic parameters. 
As expected, the best-fit $v_{COM}$ and $q$ vary greatly among different noise-realizations of the mock data, after being initialized as before.


\begin{figure*}
    \centering
    \includegraphics[width=\textwidth]{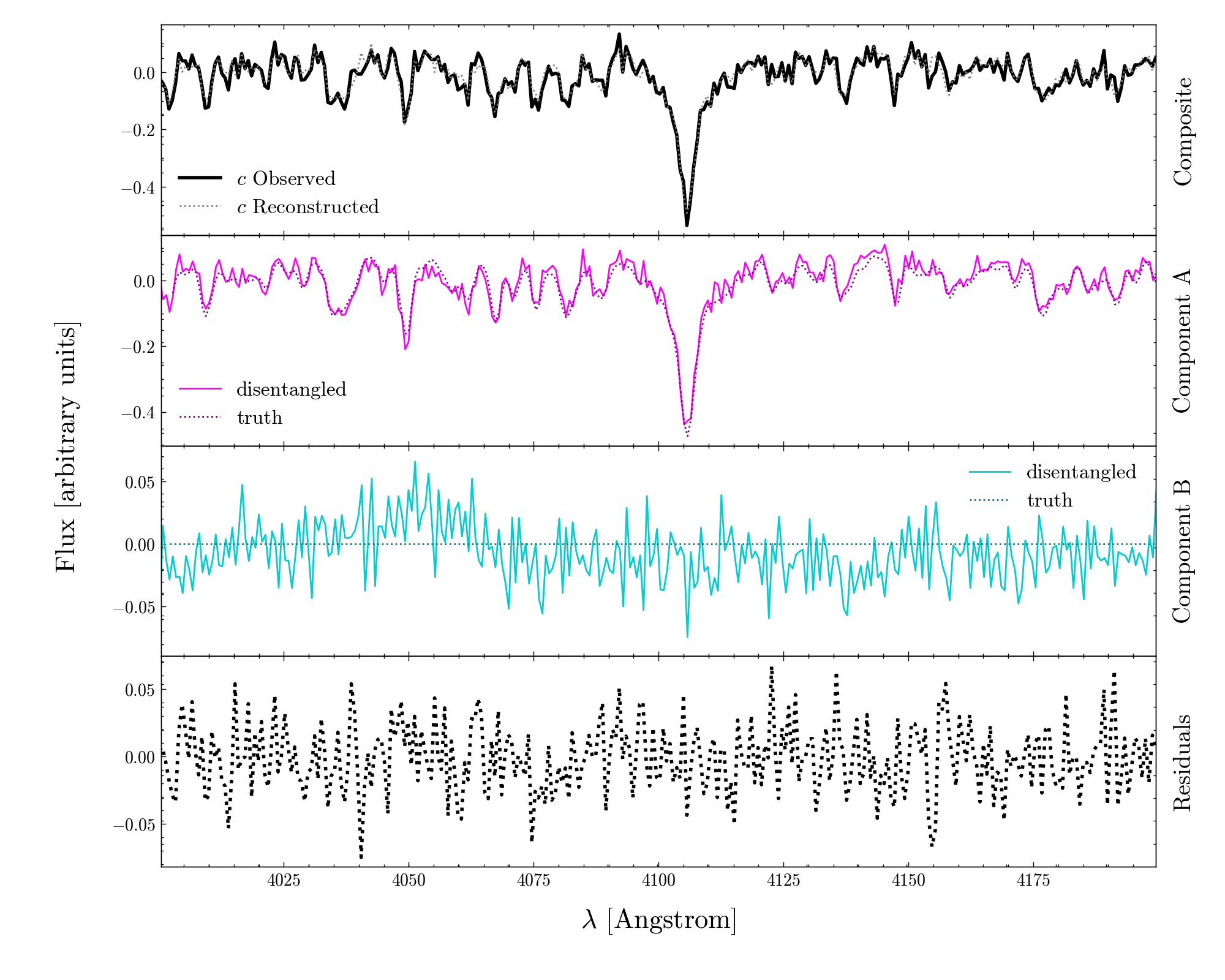}
    \caption{Formal disentangling result shown for one epoch for a simulated SB1 system, computed using the best-fit parameters found by the optimiser, analogous to figure \ref{fig:disentangled}. The spectrum of the primary is recovered, as it is the only luminous component in the system. The code also constructs of course a second component, which is close to just noise, as it should be for a dark secondary.}
    \label{fig:disentangled_a0}
\end{figure*}

However, $\alpha$ is always found to be small. This can be traced back to the fact that the disentangled spectrum of the (non-existing) ``secondary'' is close to noise, as illustrated in 
Figure \ref{fig:disentangled_a0}, which shows the disentangling results using the parameters as determined by the optimiser. The primary spectrum is of course recovered well. The computed secondary spectrum essentially appears to be noise, which then leads to the small recovered $\alpha$.

This shows that our approach degenerates gracefully for an SB1 system towards an ill-defined $v_{COM}$ and $q$, with high variance or uncertainty, and towards a small estimate of $\alpha$. Further, inspection of the actual disentangled component spectra reveals a lack of features in the secondary, in line with the input of a zero secondary.

The physically most sensible upper limit on $\alpha$ in this scenario depends on the template and its spectral features, effectively $\alpha^{max}(T_{\rm eff},$v~sin$(i)$. If one has external priors e.g. on the temperature, this can be easily incorporated into the $\alpha^{max}$ estimate.

\subsection{Sample Application: the ``Unicorn'' \& ``Giraffe''}
\label{subsec:the_unicorn_&_giraffe}

The two systems,  V723 Mon (``The Unicorn'') \& 2M04123153+6738486 (``The Giraffe''),  present excellent examples of the power of disentangling compared to cross-correlation and other template-based methods. In an initial study, \cite{jayasinghe2021unicorn} and \cite{jayasinghe2022giraffe} had reported to have found SB1 binaries with a primary mass and orbital parameters requiring the secondary to be a dormant black hole. Further study by \cite{el2022unicorns} (hereafter E22) found that the authors of the first study had been led to incorrect inferences about the nature and mass of the primary, whihc then led to an incorrectly inferred a mass limit for the secondary. In \citetalias{el2022unicorns}, spectral disentangling revealed a non-degenerate stellar secondary in both cases. Both systems were products of previously occurring mass transfer, leading to a stripped, luminous, but very low-mass primary ``masquerading'' as a much higher mass star, and a higher mass secondary. In the case of the Unicorn, the secondary had also been significantly spun up, smearing out its spectral lines.

Disentangling was able to reconstruct the component spectra and help identify the stars, but took significant attention to detail in the analysis, as well as prior knowledge of the velocities of the primary, and solid guesses about further system parameters.

In this work, we present ``blind'' disentangling of the Unicorn (with the Giraffe included in the Appendix), assuming no prior knowledge of the systems other than the assumption that they are binaries, and thus disentangling is expected to yield sensible results. We also explore the performance of the disentangler assuming the spectra had been more ``large survey style'', i.e. at lower resolution. We apply our method to the Unicorn here, as it is the more ``complicated'' of the two systems to identify, due to the very rapid rotation of the secondary.

Both in \citetalias{el2022unicorns} and in this work, we use data from the Keck/HIRES spectrograph \citep{1994SPIE.2198..362V} to identify the components of the systems. Natively, the data have resolution of $R \approx 60,000$, with 7 epochs for the Unicorn, and 8 for the Giraffe. The data cover a wavelength range from 3900 to 8000 \AA, and have a typical SNR of 20 per pixel at 5000 \AA. More details can be found in \citet{jayasinghe2021unicorn} and \citet{jayasinghe2022giraffe} for the Unicorn and Giraffe data, respectively.

\begin{figure}
    \centering
    \includegraphics[width=0.45\textwidth]{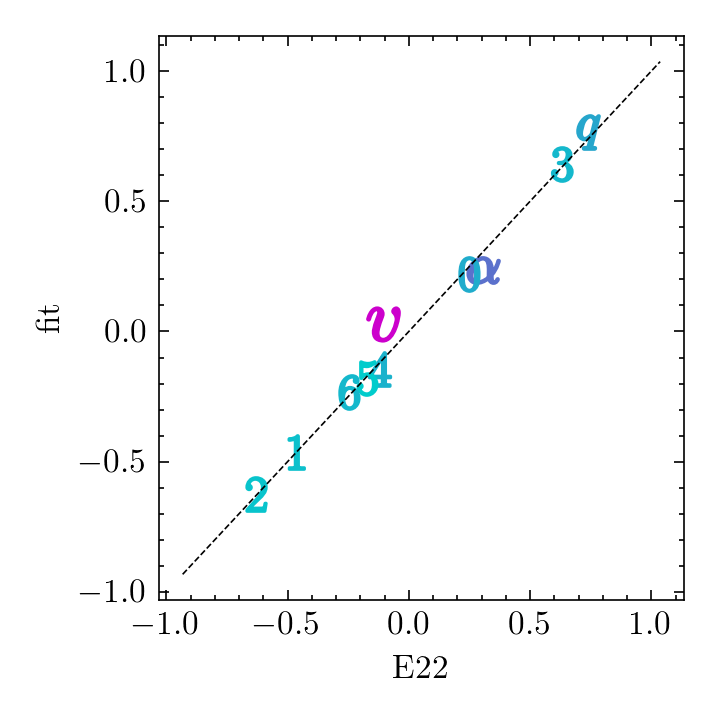}
    \caption{The parameters of the Unicorn as recovered by the optimiser (y-axis) compared to the ones found by \protect\citetalias{el2022unicorns} (x-axis). In the case of agreement, we expect the points for each parameter to lie on the grey, dashed, diagonal line.}
    \label{fig:unicorn_fom}
\end{figure}

Figure \ref{fig:unicorn_fom} shows how well the optimiser was able to recover the parameters of the Unicorn system at the native resolution of the data, $R \approx 60,000$. This was performed on the wavelength range from 5275 to 5370 \AA, as this section contains enough lines to successfully disentangle while still being narrow enough for the assumption of a constant light ratio not to break down. We see the optimiser seems to recover the individual velocities of the primary very well, owing to the clear and dominant lines in the composite spectrum. It struggles slightly more with the systemic velocity, and light ratio, likely because of the very rapid rotation and thus broadened lines of the secondary. Despite this, the light ratio is still found to an acceptable accuracy. This is likely due to a combination of the light ratio finder using both the primary and secondary spectra, and the set of templates used in the process containing non-rotating and rapidly rotating (v = 100km/s) templates.

\begin{figure}
    \centering
    \includegraphics[width=0.45\textwidth]{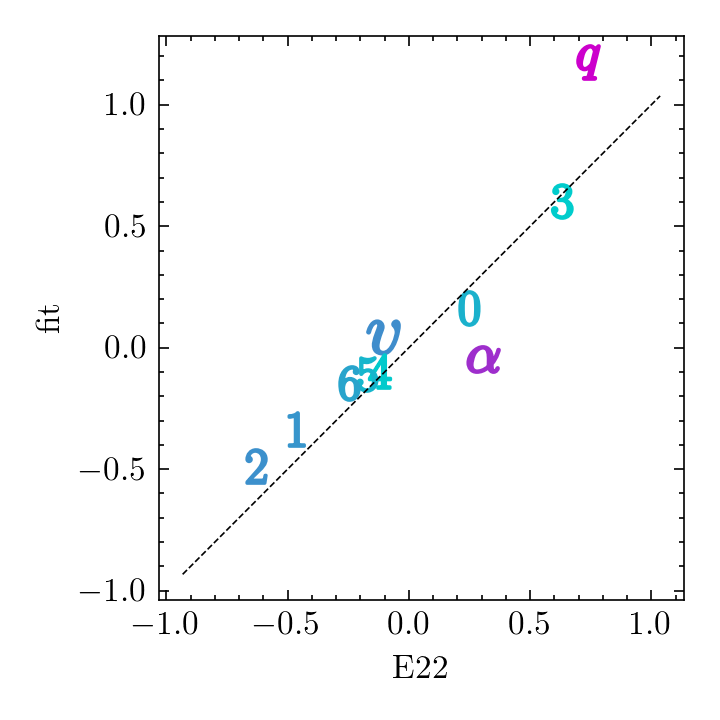}
    \caption{Analogous to Figure \ref{fig:unicorn_fom}, with an artificially reduced resolution of $R \approx 2,000$.}
    \label{fig:unicorn_fom_lowres}
\end{figure}

In contrast to this, we see Figure \ref{fig:unicorn_fom_lowres}, which shows the performance of the optimiser for the same data, now artificially resampled to a lower resolution of $R \approx 2,000$. A larger wavelength window of 5200 to 5450 \AA{} was chosen here for disentangling to account for the more broadened and thus fewer lines. We see here that the RVs tend to be underestimated, with the disentangler finding generally smaller velocities than \citetalias{el2022unicorns}. This is unsurprising, as the lower resolution leads to broader CC peaks, where eventually the peaks of the primary and secondary velocities are so broad that the secondary peak ``drags'' the primary down. Because of this underestimation in velocities and the broadened lines of the secondary, the optimiser also cannot correctly assess the mass ratio. 
However, we do see that the optimiser has pushed q away from its initial starting guess of $log(q) = 0$ towards higher values, meaning the data do tell us that the secondary is heavier than the primary, but struggles to perfectly determine how much heavier. As a consequence of this, the disentangled spectra are less accurate, also making it difficult to find the correct light ratio.

\begin{figure*}
    \centering
    \includegraphics[width=\textwidth]{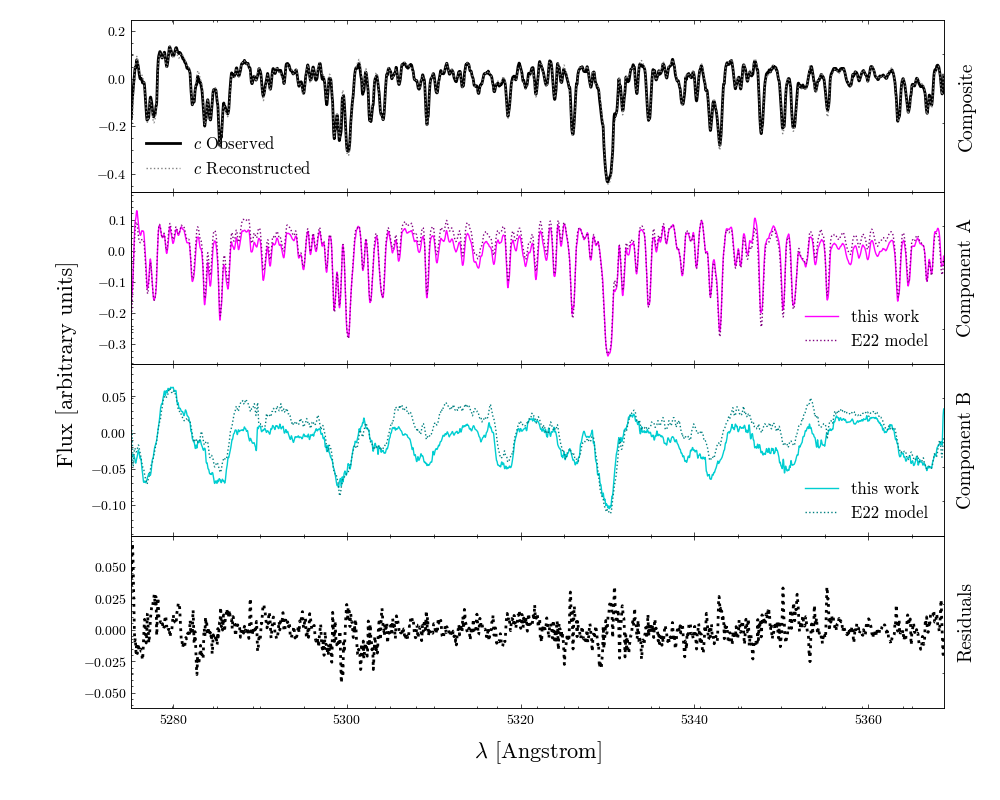}
    \caption{The results of the autonomous disentangling applied to the Unicorn for one epoch. The top panel shows the observed spectrum (black, solid line) as well as the reconstruction from the velocity shifted disentangled component spectra (grey, dashed line). The other two panels display the disentangled rest-frame solutions for the primary and secondary, respectively, in magenta and cyan. The dashed, darker lines in these two panels indicate the model spectrum found by \protect\citetalias{el2022unicorns} to most closely fit the disentangled solutions found there. The bottom panel shows the residual between the reconstructed and observed composite spectra. We see generally good agreement between the solutions from this work and \protect\citetalias{el2022unicorns} models.}
    \label{fig:unicorn}
\end{figure*}

In Figure \ref{fig:unicorn}, we see the disentangled Unicorn spectra for the native resolution of $R \approx 60,000$, as well as the models from \citetalias{el2022unicorns}. Here, we see generally good agreement between the two, with one of the more apparent differences stemming from the slightly different light ratios between their work and this one. For example, we see that the \citetalias{el2022unicorns} model for the secondary generally shows slightly ``stronger'' lines, i.e. a bigger difference between maxima and minima. There does not seem to be a significant offset along the wavelength axis, suggesting that the systemic velocity was recovered (mostly) accurately, at least to within the resolution limit. 

\begin{figure*}
    \centering
    \includegraphics[width=\textwidth]{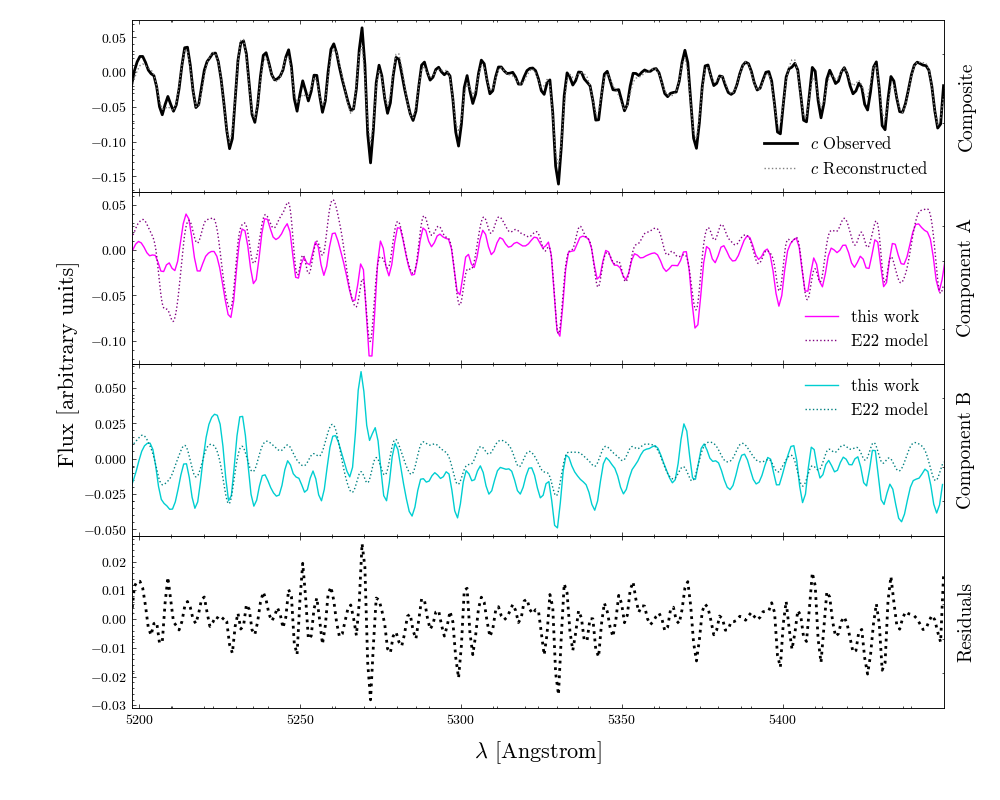}
    \caption{Analogous to Figure \ref{fig:unicorn}, with an artificially reduced resolution of $R \approx 2,000$.}
    \label{fig:unicorn_lowres}
\end{figure*}

Figure \ref{fig:unicorn_lowres} shows the disentangling result for the $R \approx 2,000$ case, zoomed out more to cover a larger wavelength range than Figure \ref{fig:unicorn}. We elected to do this due to the ``smearing out'' at lower resolutions that removes many of the finer features and broadens the dominant lines. To still show a good number of features, a wider wavelength window is necessary. Despite the underestimations of the velocities and mass ratio, the disentangled component spectra still resemble the (downsampled) models from \citetalias{el2022unicorns}. Many of the visible features are recovered, even though the underestimated velocities likely lead to some of the features of the primary being more ``washed out''. Similarly, these in conjunction with the incorrect $q$ have an analogous effect on the recovered spectrum of the secondary. The issues with the light ratio then arise naturally as a consequence of the recovered spectra being more ``smeared out'' than the templates we are comparing to. Additionally, for larger wavelength window, the assumption of a constant light ratio across the spectrum holds less true, which is likely a contributing factor here.

\section{Discussion and Summary}
\label{sec:discussion_and_summary}

In this work, we have set out to implement spectral disentangling as described by \cite{simon1994disentangling} and \cite{hadrava1995orbital}, among others, in an algorithm tailored for million-stars surveys. This meant adapting the process to be suitable for the regime of few epochs, modest S/N, and moderate resolution ($R \approx 2,000$). At the same time, we required the code to be robust in a range of scenarios (e.g. both hot and cool stars with a range of companions), as well as able to autonomously find the parameters necessary for the disentangling procedure precisely and accurately. Given the goal of applying it to large volumes of data, the code also needed to be fast.

Our approach has been to combine a number of known aspects and strategies to create an end-to-end autonomous pipeline that fulfills all of the above criteria, while consisting of a number of internal steps. We have implemented wavelength-space disentangling based on \cite{simon1994disentangling}; Tikhonov regularisation \citep{tikhonov1963solution, phillips1962technique} to ensure smoothness of the disentangled spectra; downhill-simplex optimisation \citep{nelder1965simplex} to optimise parameters; template cross-correlation and TIRAVEL \citep{zucker2006tiravel} to get initial velocity guesses; as well as template-based methods to find the systemic velocity and light ratio of the system, all in one. We have explored the efficacy of this process using a variety of synthetic SB2 spectra that emulate those provided by the large surveys we developed this pipeline for.

We have found overall satisfactory performance on our simulated systems, as well as a real-life example (the Unicorn), which had previously been misidentified \citep{jayasinghe2021unicorn}, and required great care to analyse correctly \citepalias{el2022unicorns}. We have shown that in the target regime of few epochs, modest S/N and moderate resolution, the algorithm can still recover the relevant system parameters and the component spectra for many of the systems explored. However, there are still areas of the parameterspace where our method fails or encounters issues:

\begin{itemize}
    \item Extreme light ratio, relative to S/N. For light ratios below $\alpha \sim 0.1$, while the algorithm generally still recovers the primary and associated velocities correctly, the spectrum of the secondary is overwhelmed by the noise, and thus both the spectrum of the secondary and the mass ratio $q$ cannot be determined correctly. In this case, the disentangler reacts similarly as if the system were an SB1, meaning we can set a light ratio ``cut-off'', below which we can no longer separate an SB1 from an SB2.
    \item Low velocity semi-amplitude, relative to resolution. If the velocity semi-amplitude (especially of the primary) falls close to or below the minimum resolvable velocity, the first step of the optimisation, finding starting guesses for the velocities of the primary, fails, and subsequent steps cannot rectify the issue. Here, the majority of parameters are determined incorrectly, and thus the disentangled spectra are also spurious.
    \item Equal-mass binaries. For systems where the primary and secondary are of equal mass, or close to it, the associated spectra look too similar, introducing an ambiguity as to which set of lines belongs to which object at which epoch. In this case, the velocities for the primary and secondary are degenerate for any one epoch (as the method cannot tell ``which is which''), causing further issues with the disentangler. One method to rectify this might be to invoke orbital fitting, however this would only be possible in the regime of sufficient epochs. Careful, individual treatment of these systems might provide another path to a successful solution, but goes against the spirit of this work, and autonomy of the process. However, despite the issues with finding the correct parameters, the disentangled spectra often still resemble the truth quite well. This is because a simple switch of the velocities in the matrix solver step simply means that the spectra are swapped for the affected epoch. If the spectra are the same or very similar, this has only a minor effect on the disentangling result. 
    \item Hot stars. Due to the relatively few and broad lines of hot stars compared to cooler stars, the disentangler encounters more issues for these objects. The broader lines lead to a less precise determination of radial velocities, which creates problems when trying to find the mass ratio, spectra, and other parameters. In many cases, the disentangler does still arrive at a satisfactory solution, but less consistently than for a similar system in terms of light ratio and velocity semi-amplitude but with a cooler primary.
\end{itemize}

One of the novelties of this work is the inclusion of Tikhonov regularisation \citep{tikhonov1963solution, phillips1962technique} to ensure that spectra are smooth. This process allows us to remove much of the high-frequency noise from the solution, much like a truncated SVD would, but it is independent of the solver method employed and allows us to use fast, iterative algorithms to solve for the spectra while retaining desired characteristics. It also does not enforce physical constraints on the solutions, allowing a great deal of ``freedom'' when determining the component spectra.


An important aspect of survey disentangling is the sheer volume of data to be processed, which requires short optimisation times for any one object and consistent optimisation. We have parallelised the code using Python's \verb|multiprocessing| library. On 72 CPUs, the computation of the parameters and spectra of 120 simulated systems (2 masses, 2 resolution regimes, 5 different light ratios, and 6 different velocity semi-amplitudes, see section \ref{subsec:simulated_data} for details) took about 3 minutes. This includes running each system through the optimiser 6 times (once for each epoch that is removed from the data) for bootstrapping. As the systems are all independent of each other, the process scales without much difficulty to larger clusters with more cores.  This bodes well for the application to even a million stellar systems.

The examples provided in this work have demonstrated that even in the regime of few epochs, moderate resolution, and signal-to-noise ratio, disentangling poses a very viable option for analyzing spectroscopic binaries. This, as well as the fast runtime of the algorithm, allows disentangling to be performed on large surveys, such as LAMOST \citep{cui2012large} and SDSS-V \citep{york2000sloan}. The SDSS-V catalog will contain multi-epoch spectra of $\sim 200,000$ hot and massive stars, which are predicted \citep{sana2012binary, moe2017mind} to have a high multiple fraction, making them great candidates for disentangling. Thus, the method developed here will allow us to weed out contaminants in our continued search for dark companions, and select ideal targets for higher-resolution, better-S/N and more-epoch follow up investigation.

\section*{Acknowledgements}
\label{sec:acknowledgements}

RS and HWR acknowledge the European Research Council for support from the ERC Advanced Grant ERC-2021-ADG101054731.

We thank Tsevi Mazeh, Silvia Almada Monter, Johanna Müller-Horn and Jaime Villaseñor for stimulating and helpful discussion.

We also thank Tomer Shenar for insightful and constructive comments on the manuscript.

This publication made extensive use of the online authoring Overleaf platform (\url{https://www.overleaf.com/}).\\
The data processing and analysis made use of
        matplotlib \citep{Hunter:2007},
        NumPy \citep{harris2020array},
        the IPython package \citep{PER-GRA:2007},
        SciPy \citep{2020SciPy-NMeth},
        AstroPy \citep{astropy:2013, astropy:2018, astropy:2022}
        SpectRes \citep{carnall2017spectres}
        and SpectResC \citep{lam_2023_7879105}

\section*{Data Availability}
\label{sec:data_availability}
 
Data used in this study are available from the corresponding author upon request.



\bibliographystyle{mnras}
\bibliography{mnras_template} 




\appendix

\section{Method - Further Detail}
\label{sec:method-detail}

\subsection{Normalisation} 

Spectra are generally normalised before analysis is performed. 
Here, we seek to perform continuum normalisation, where each datapoint is divided by the continuum value at its position. Since the distance from a star is comparatively hard to determine accurately, it is advantageous to remove it as a fitting parameter from the problem. Continuum normalisation aids in doing that, by removing information about the specific flux at each pixel, while still retaining the important information contained in the line depth relative to the continuum.


Generally, for continuum normalisation, a polynomial or spline is fit to the spectrum, representing the low-frequency variations, and then subtracted. In this work, each datapoint is divided by the median value of $Q$ pixels around it, where the value of $Q$ governs what this normalisation looks like. $Q$ values close to $N_{px}$ will lead to a normalised spectrum that mostly keeps its shape, but now has an average flux near 1. $Q$ values close to 1 will lead to a flattening out of the entire spectrum, and a loss of features; it is thus important to choose the size of $Q$ correctly. More specifically, the window needs to be wide enough to fully cover the wider lines in the spectrum as not to create ``humps'' or ``wings'' on either side when normalising, but narrow enough to efficiently remove the shape of the continuum. Further, when normalising in this manner, we wish to ignore pixels with associated NaN flux values.

Finally, we subtract 1 from the normalised spectrum. This leads to an average flux near 0, with emission lines extending above, and absorption lines below 0. We do this to effectively remove low-frequency variations from the disentangling process completely, and allow the lines to dominate the process.

\subsection{Interpolation}
\label{subsec:interpolation}

As in the disentangling prescription used in this work, the shifts $\Delta \Lambda$ are not necessarily of integer pixel value, we need a way to evaluate the component spectra at intermediate pixel values; for one component, this can be achieved by linear interpolation as follows:

\begin{equation}
     x(\Lambda) = r_j \cdot x(\Lambda_{n+1}) + (1-r_j) \cdot x(\Lambda_{n}),
\end{equation}

where $\Lambda$ is some intermediate wavelength between $\Lambda_n$ and $\Lambda_{n+1}$, and $r_j$ a shift-dependent (and thus epoch-dependent) interpolation factor given by

\begin{equation}
    r_j = \frac{\Lambda - \Lambda_{n}}{\Lambda_{n+1} - \Lambda_n}.
\end{equation}

Here, the numerator represents the distance of the wavelength at which we want to evaluate, $\Lambda$, from the nearest lower integer pixel wavelength $\Lambda_{n}$, and the bottom represents the distance between two integer pixel wavelengths. $r_j$ thus gives the fractional distance between $\Lambda$ and $\Lambda_n$. The relative contribution of the spectrum at $\Lambda_n$ to the linearly interpolated spectrum at $\Lambda$ is then $1-r_j$, the complement of $r_j$ to unity. Thus, the fractional contribution of the spectrum at $\Lambda_{n+1}$ to the linear interpolation at $\Lambda$ is $r_j$. 

This makes intuitive sense: if the interpolated point is close to $\Lambda_{n}$, $r_j$ is close to 0, and most of the contribution to $x(\Lambda)$ comes from $x(\Lambda_{n})$, thus its associated coefficient ($1-r_j$) is close to 1, while $x(\Lambda_{n+1})$'s coefficient ($r_j$) is close to 0. Conversely, if the interpolated point is close to $\Lambda_{n+1}$, $r_j$ is close to 1, and most of the contribution to $x(\Lambda)$ comes from $x(\Lambda_{n+1})$, thus its associated coefficient ($r_j$) is close to 1, while $x(\Lambda_{n})$'s coefficient ($1-r_j$) is close to 0.

At this point, it should be noted that this also needs to work if the shift is larger than 1 pixel. For shifts between -1 and 1 pixels, $\Lambda_n$ and $\Lambda_{n+1}$ are simply the original pixel, and the one to the left/right of it. For larger shifts, the pixels need to be chosen accordingly, such that they ``bracket'' the intermediate pixel. 

This linear interpolation can be generalised to higher orders, using a lagrange polynomial:

\begin{equation}
x(\Lambda)=\sum_{o=0}^{O} w_{o}(\Lambda) \cdot x\left(\Lambda_{o^{\prime}}\right),
\end{equation}

where O is the order of the polynomial,

\begin{equation}
o^{\prime}=o-f^K_{j}-\rm{int}\left(\frac{O}{2}\right).
\end{equation}

Here, $f^K_{j}$ is the result of the floor operation on the wavelength shift in pixels of component k at epoch j. Then, w can be computed from:

\begin{equation}
w_{o}(\Lambda)=\prod_{\substack{0 \leq m \leq O \\ m \neq o}} \frac{\Lambda-\Lambda_{m^{\prime}}}{\Lambda_{o^{\prime}}-\Lambda_{m^{\prime}}},
\end{equation}

with 

\begin{equation}
m^{\prime}=m-f^K_{j}-\rm{int} \left(\frac{O}{2}\right).
\end{equation}

In the case of $O=1$, this reduces to the linear case discussed above.

Using the interpolation, Eq. \ref{prematrix} can then be written as:

\begin{multline}
    \vec{c}_{j, pred} = \left( \{r^A_{j} \cdot x^A(\Lambda^A_{j,n+1}) + (1-r^A_{j}) \cdot x^A(\Lambda^A_{j,n})\right) \\
    + \left( r^B_{j} \cdot x^B(\Lambda^B_{j,n+1}) + (1-r^B_{j}) \cdot x^B(\Lambda^B_{j,n})\}\right).
    \label{prematrix2}
\end{multline}

\subsection{Matrix Structure: Details}
\label{subsec:matrix_structure}

\label{app:matrix}

The index of the first non-zero element in the first row of each block of $\mathbf{M}$ depends on the size of the shift. 

Let $s^K_{j}$ be the shift of the spectrum of component $k$ at epoch $j$ in units of the resolution element $\delta$:

\begin{equation}
    s^K_{j} = \frac{\Delta \Lambda^K_{j}}{\delta}.
\end{equation} 

We then make use of the $floor(m)$ operation, which returns the next lowest integer for any number $m$; for example, $floor(3.2)$ returns $3$, $floor(-1.5)$ returns $-2$, $floor(0.7)$ returns $0$, etc. For component $k$ at epoch $j$, then:

\begin{align}
    f^K_{j} &\equiv floor(s^K_{j})\\
    r^K_{j} &\equiv s^K_{j} - f^K_{j}.
    \label{nicernot}
\end{align}

Here, $r^K_{j}$ gives distance in the positive $\Lambda$ direction from $f^K_{j}$ for each $s^K_{j}$ - specifically, this $r^K_{j}$ is equal 
to the $r^A_{j}$ or $r^B_{j}$ in Equation~\ref{prematrix2}.
$f^K_{j}$, then, is the nearest lower integer pixel for each shift $s^K_{j}$, 
meaning $\Lambda_{i+ f^K_{j}}$ is equivalent to $\Lambda^A_{j,n}$ or $\Lambda^B_{j,n}$ in Equation \ref{prematrix2}. It gives the index of the first non-zero element of the first row of each block of $\mathbf{M}$, or, equivalently, the shift of the (off-)diagonal. 
The values of this (off)diagonal are $1-r^K_{j}$. Consequently, $\Lambda_{i+ f^K_{j}+1}$ ($\equiv \Lambda^A_{j,n+1}$ or $\Lambda^B_{j,n+1}$ in Equation \ref{prematrix2}) gives the index of the second non-zero element of the first row of each block of the matrix, or the shift of its associated (off-)diagonal, with values $r^K_{j}$. 

For a shift in the negative $\Lambda$ direction that is smaller than 1 integer pixel, the block for component $k$ and epoch $j$ would look as follows:

\NiceMatrixOptions{cell-space-limits = 1pt}

\begin{center}
$
\begin{pNiceMatrix}
1-r^K_{j} & r^K_{j} & 0 & 0 & 0 & ... \\
0 & 1-r^K_{j} & r^K_{j} & 0 & 0 &  ...\\
0 & 0 & 1-r^K_{j} & r^K_{j} & 0 &  ...\\
0 & 0 & 0 & 1-r^K_{j} & r^K_{j} &   ...\\
... & ... & ... & ... & ... & ...\\
\end{pNiceMatrix}$
\end{center}

For a shift in the negative $\Lambda$ direction that is between 1 and 2 integer pixels, the block for component $k$ and epoch $j$ would look as follows:

\begin{center}
$
\begin{pNiceMatrix}
0 & 1-r^K_{j} & r^K_{j} & 0 & 0 &  ... \\
0 & 0 & 1-r^K_{j} & r^K_{j} & 0 &   ...\\
0 & 0 & 0 & 1-r^K_{j} & r^K_{j} &   ...\\
0 & 0 & 0 & 0 & 1-r^K_{j} &    ...\\
... & ... & ... & ... & ... & ...\\
\end{pNiceMatrix}$
\end{center}

For a shift in the positive $\Lambda$ direction that is smaller than 1 integer pixel, the block for component $k$ and epoch $j$ would look as follows:
\begin{center}
$
\begin{pNiceMatrix}
r^K_{j} & 0 & 0 & 0 & 0 & ... \\
1-r^K_{j} & r^K_{j} & 0 & 0 & 0 &  ...\\
0 & 1-r^K_{j} & r^K_{j} & 0 & 0 &  ...\\
0 & 0 & 1-r^K_{j} & r^K_{j} & 0 &   ...\\
... & ... & ... & ... & ... & ...\\
\end{pNiceMatrix}$
\end{center}

The task of disentangling the composite spectra from all epochs into two rest-frame spectra can then be expressed as a minimisation of the sum over epochs:

\begin{equation}
    \vec{x}_{best} = \argmin_{\vec{x}^A, \vec{x}^B} \sum_j^{N_{ep}}\lVert  \vec{x}^A_{j} + \vec{x}^B_{j} - \vec{c}_j\rVert_2.
\end{equation}

This can be written, more explicitly, as the sum over epochs and pixels: 
\begin{multline}
    \vec{x}_{best} = \argmin_{\vec{x}^A, \vec{x}^B} \sum_j^{N_{ep}} \sum_i^{N_{px}} \lVert  x^A(\Lambda^A_{j,i} - \Delta \Lambda^A_{j}) \\ + x^B(\Lambda^A_{j,i} - \Delta \Lambda^B_{j}) - c_j(\Lambda_i)\rVert_2.
\end{multline}

Then using the interpolation described, as well as the notation in Equation \ref{nicernot}, we get:
\begin{multline}
    \argmin_{\vec{x}^A, \vec{x}^B} \sum_j^{N_{ep}} \sum_i^{N_{px}} \lVert ( r^A_{j} \cdot x^A(\Lambda_{i+ f^A_{j}+1} ) \\ 
     + (1-r^A_{j}) \cdot x^A(\Lambda_{i + f^A_{j}} )\\
     + r^B_{j} \cdot x^B(\Lambda_{i + f^B_{j}+1} ) \\
     + (1-r^B_{j}) \cdot x^B(\Lambda_{i + f^B_{j}} ) - c_j(\Lambda_i) \rVert_2.
\end{multline}

\subsection{Selection and scaling of optimisation parameters}
\label{subsec:optimisation_parameters}

For most optimisers it is advantageous to have normalised parameters. In this case, we are performing some rescaling by taking the common logarithm of $q$, and dividing all epoch velocities ($\vec{v}^A$) by 100. This ensures that the optimiser can take similarly large steps in different directions with relative ease, as well as changing the mass ratio to a more sensible scale - a step from a $q = 1$ to  $q = 2$ is relatively large, meaning a doubling of the secondary mass, while a step from $q = 50$ to $q = 51$ will cause very little difference in the overall solution. In log-space however, steps of similar size have a more consistent effect along the parameter range.

Even with these considerations finding these parameters is not trivial, considering the dimensionality of the parameter space is frequently high. \citep{simon1994disentangling}, \cite{hadrava1995orbital}, \cite{sablowski2017spectral} and others have suggested optimising over the orbital parameters instead, which can reduce the dimensionality of the problem to 7, which, in the case of many epochs, can be significantly less than fitting each epoch velocity individually. Here, however, we wish to demonstrate that even for a smaller number of epochs, spectral disentangling is viable, and thus optimise on the velocities and mass ratio.

\subsection{TIRAVEL}
\label{subsec:tiravel}

If we wish to forgo using templates to find initial guesses for the velocities of the primary, or do not have suitable templates available, we can employ the TIRAVEL algorithm \citep{zucker2006tiravel} in order to find the relative inter-epoch velocities. This algorithm is most suitable to SB1s, or SB2s with comparatively faint secondaries. TIRAVEL achieves this by computing the cross-correlations between all epochs and then determining the velocitiy vector, with an offset, that solves for all the inter-epoch velocities by maximising the value of the CC at the inter-epoch velocity shifts. The offset arises from the fact that this method employs no templates, and thus no knowledge of what a rest-frame spectrum looks like is included. In this implementation, we set the observed RV of the first epoch to 0, and then calculate the other RVs relative to this. Once TIRAVEL has computed a candidate velocity vector, we create an estimation of the primary's spectrum by shifting all epoch spectra by their relative RVs and co-adding the spectra. This should, in theory, minimise the contribution from the secondary (in an SB2) and boost the signal of the primary - being more efficient the more epochs are available. 

The gross offset of all velocities from their true values might seem problematic at first, but we remind ourselves that the systemic velocity is corrected for in a later step in our disentangling pipeline. This also accounts for any constant offset among all velocities that arises due to TIRAVEL, and thus this should cause no further issues.

\section{Additional Results}
\label{sec:additional_results}

\subsection{High Resolution Simulations}
\label{subsec:highres_simulations}

We repeat the simulations from section \ref{sec:algorithm_validation} for the same systems, but now with a higher resolution of $R \approx 20,000$. Here, the disentangler's difficulty with (close to) equal mass binaries becomes even more apparent, demonstrating that this is not a resolution issue. We also note that there are now fewer problems arising for small velocities (such as for the K=50 km/s systems), as the resolving power is now high enough to accurately determine even the smaller velocity shifts. Further, we see that the higher mass (3 $M_{odot}$) primary still generally poses a bigger challenge than the lower mass (1.1 $M_{odot}$) primary, again, due to the wider lines and fewer features overall.

\begin{figure*}
    \centering
    \includegraphics[width=\textwidth]{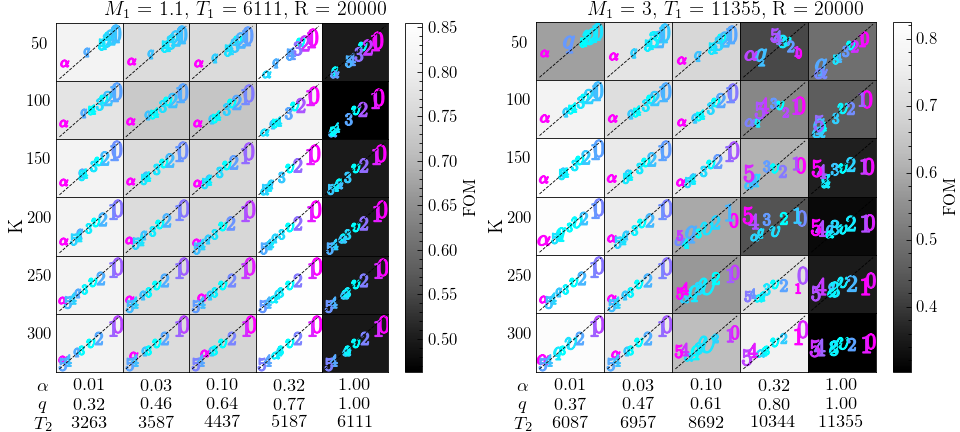}
    \caption{As Figure \ref{fig:grid}, with a higher resolution of $R \approx 20,000$. }
    \label{fig:grid_hires}
\end{figure*}

\begin{figure*}
    \centering
    \includegraphics[width=\textwidth]{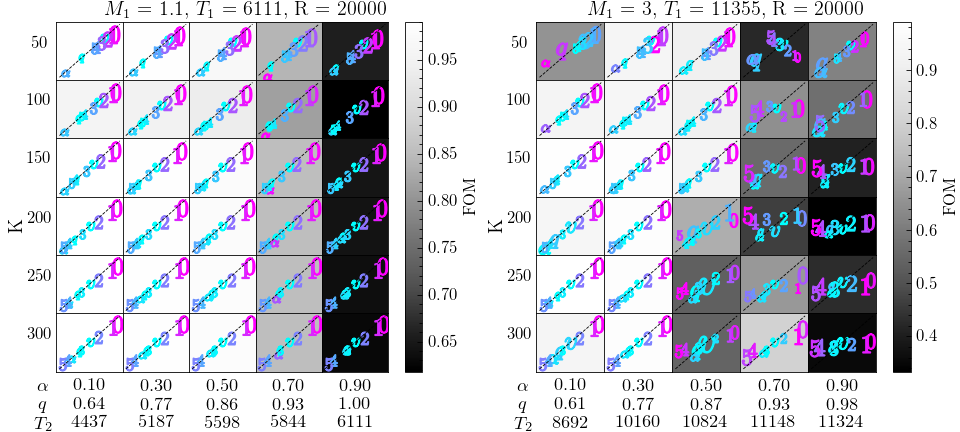}
    \caption{As Figure \ref{fig:grid_small}, with a higher resolution of $R \approx 20,000$. }
    \label{fig:grid_small_hires}
\end{figure*}

\subsection{The ``Giraffe''}
\label{subsec:giraffe}

We perform ``blind'' disentangling on the Giraffe system, both in the native resolution regime of $R \approx 60,000$, and on spectra artificially reduced to $R \approx 2,000$, as with the Unicorn in section \ref{subsec:the_unicorn_&_giraffe}. 

\begin{figure}
    \centering
    \includegraphics[width=0.45\textwidth]{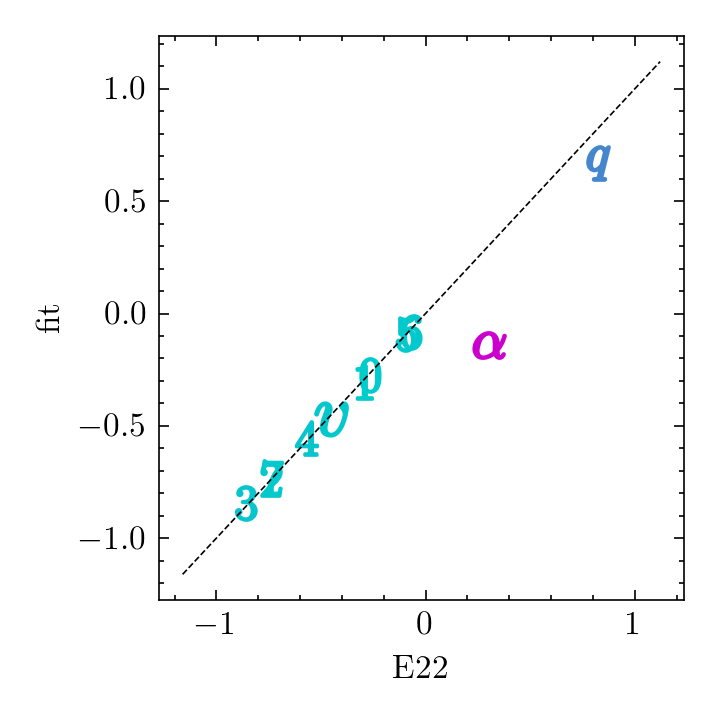}
    \caption{The parameters of the giraffe as recovered by the optimiser (y-axis) compared to the ones found by \protect\citetalias{el2022unicorns} (x-axis). in the case of agreement, we expect the points for each parameter to lie on the grey, dashed, diagonal line.}
    \label{fig:giraffe_fom}
\end{figure}

\begin{figure*}
    \centering
    \includegraphics[width=\textwidth]{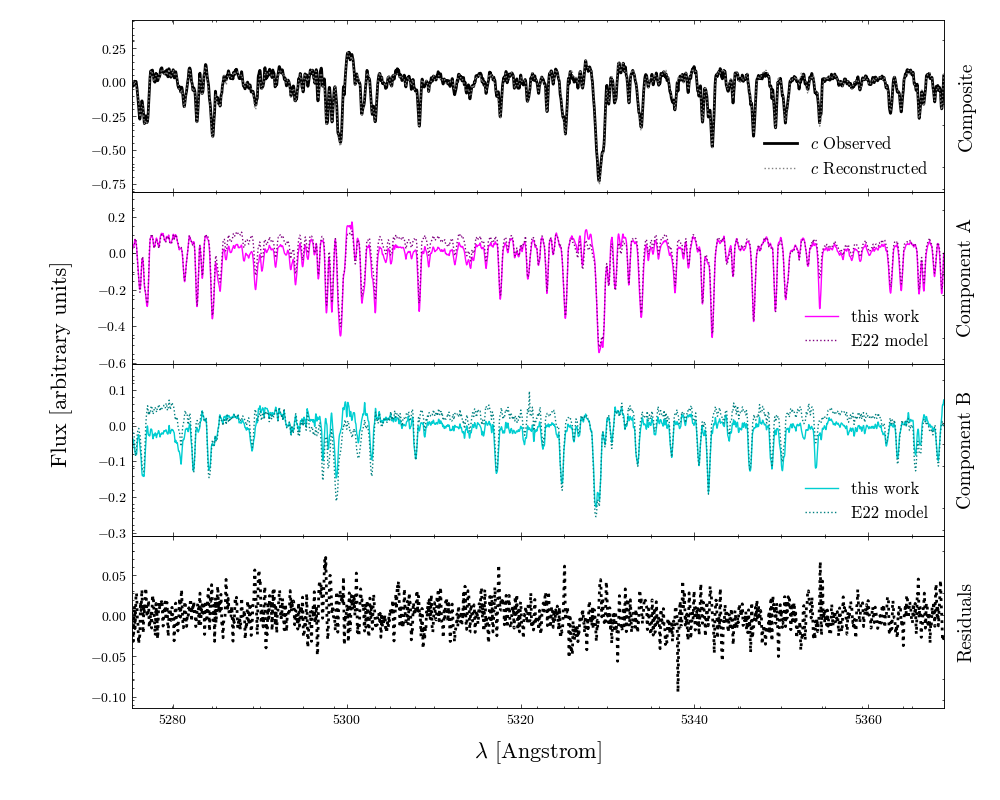}
    \caption{The results of the autonomous disentangling applied to the Giraffe for one epoch. The top panel shows the observed spectrum (black, solid line) as well as the reconstruction from the velocity shifted disentangled component spectra (grey, dashed line). The other two panels display the disentangled rest-frame solutions for the primary and secondary, respectively, in magenta and cyan. The dashed, darker lines in these two panels indicate the model spectrum found by \protect\citetalias{el2022unicorns} to most closely fit the disentangled solutions found there. The bottom panel shows the residual between the reconstructed and observed composite spectra. We see generally good agreement between the solutions from this work and \protect\citetalias{el2022unicorns} models.}
    \label{fig:giraffe}
\end{figure*}

In Figure \ref{fig:giraffe_fom} we see the performance of the optimiser on the parameters of the Giraffe in the higher resolution regime, with the wavelength window the same as for the Unicorn. Compared to the Unicorn, we note a markedly higher difficulty in recovering the light ratio correctly. This is likely due to the fact that the primary and secondary spectra of the Giraffe look more similar to each other than those of the Unicorn. Due to the rapid rotation of the Unicorn's secondary, the lines of that component are significantly broadened, leading to a somewhat non-typical spectrum that is quite distinct in shape from the spectrum of the primary. Here, however, the secondary is rotating much more slowly, thus this effect is minimal. Looking at Figure \ref{fig:giraffe} lends credence to this idea, as we can see the two spectra are fairly similar in shape. We also see the effect of the underestimation of the light ratio: relative to the \citetalias{el2022unicorns} models, the lines of the primary spectrum appear deeper (meaning a larger fraction of total light is attributed to the primary, and thus the light ratio is lower).

\begin{figure}
    \centering
    \includegraphics[width=0.45\textwidth]{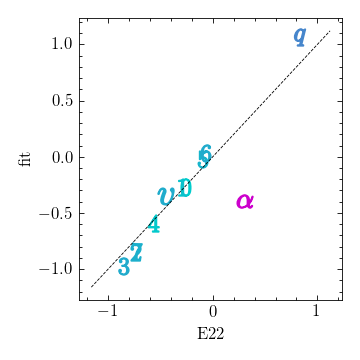}
    \caption{Analogous to Figure \ref{fig:giraffe_fom}, with an artificially reduced resolution of $R \approx 2,000$.}
    \label{fig:giraffe_fom_lowres}
\end{figure}

\begin{figure*}
    \centering
    \includegraphics[width=\textwidth]{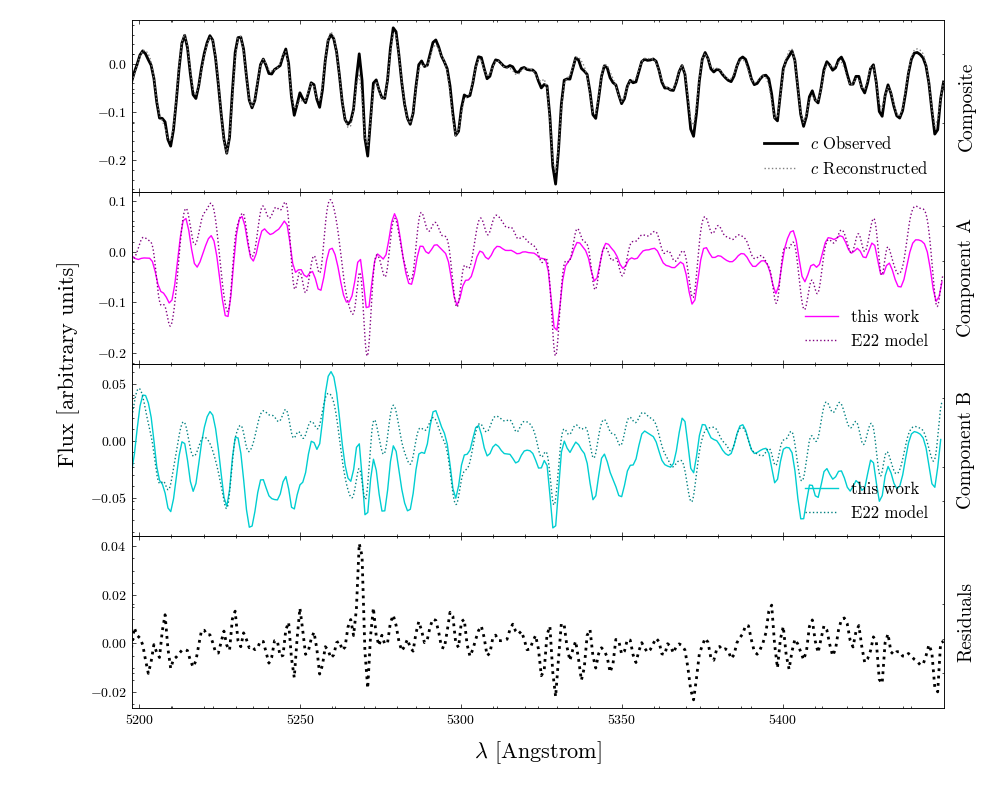}
    \caption{Analogous to Figure \ref{fig:giraffe}, with an artificially reduced resolution of $R \approx 2,000$.}
    \label{fig:giraffe_lowres}
\end{figure*}

Figure \ref{fig:giraffe_fom_lowres} shows the performance of the optimiser on the data at the artificially reduced resolution, again on a larger wavelength window analogous to the Unicorn. We note the same difficulty with accurately assessing the light ratio, as well as some small issues with the velocities, and, resulting from this, the mass ratio, owing to the lower resolving power. The effects of the incorrect light ratio in particular become very apparent in Figure \ref{fig:giraffe_lowres}. Many lines of the primary appear deeper in the \citetalias{el2022unicorns} models than in the spectrum found here, due to the too-low light ratio found by the optimiser.

\bsp	
\label{lastpage}
\end{document}